\documentclass[a4paper,fleqn]{cas-sc}

\usepackage[numbers]{natbib}

\usepackage{xcolor}           
\usepackage{tikz}
\usetikzlibrary{positioning, arrows.meta, shapes.geometric, patterns, calc}
\usepackage{changepage}

\usepackage{tabularx}
\usepackage{booktabs}
\usepackage{array}
\usepackage{threeparttable}
\usepackage{caption}

\usepackage{amsmath}
\usepackage{siunitx}
\usepackage{enumitem}
\usepackage{pifont}
\newcommand{\cmark}{\ding{51}}
\newcommand{\ccmark}{\cmark\cmark}

\usepackage{listings}
\usepackage{etoolbox}
\robustify{\ding}

\usepackage{graphicx}
\usepackage{placeins}
\usepackage{float}

\def\tsc#1{\csdef{#1}{\textsc{\lowercase{#1}}\xspace}}
\tsc{WGM}
\tsc{QE}
\tsc{EP}
\tsc{PMS}
\tsc{BEC}
\tsc{DE}

\begin{document}
\let\WriteBookmarks\relax
\def\floatpagepagefraction{1}
\def\textpagefraction{.001}

\shorttitle{RAE-CRS framework}

\shortauthors{R. Mahmud et~al.}

\title [mode = title] {Recommendation-as-Experience: A framework for context-sensitive adaptation in conversational recommender systems}



%

\author[1]{Raj Mahmud}
[orcid=0000-0002-0467-5246]
\cormark[1]


\ead{raj.pgm@outlook.com}


\affiliation[1]{organization={University of Technology Sydney, School of Computer Science},
    country={Australia}}

\author [2]
{Shlomo Berkovsky}

\affiliation[2]{organization={Macquarie University, Institute of Health Innovation},
    country={Australia}}

\author [1] {Mukesh Prasad}

\author [1] {A. Baki Kocaballi}

\cortext[cor1]{Corresponding author}



\begin{abstract}
While Conversational Recommender Systems (CRS) have matured technically, they frequently lack principled methods for encoding latent experiential aims as adaptive state variables. Consequently, contemporary architectures often prioritise ranking accuracy at the expense of nuanced, context-sensitive interaction behaviours. This paper addresses this gap through a comprehensive multi-domain study ($N = 168$) that quantifies the joint prioritisation of three critical interaction aims: \textit{educative} (to inform and justify), \textit{explorative} (to diversify and inspire), and \textit{affective} (to align emotionally and socially). Utilising Bayesian hierarchical ordinal regression, we establish domain profiles and perceived item value as systematic modulators of these priorities. Furthermore, we identify stable user-level preferences for autonomy that persist across distinct interactional goals, suggesting that agency is a fundamental requirement of the conversational experience. Drawing on these empirical foundations, we formalise the \textit{Recommendation-as-Experience} (RAE) adaptation framework. RAE systematically encodes contextual and individual signals into structured state representations, mapping them to experience-aligned dialogue policies realised through retrieval diversification, heuristic logic, or Large Language Model based controllable generation. As an architecture-agnostic blueprint, RAE facilitates the design of context-sensitive CRS that effectively balance experiential quality with predictive performance.
\end{abstract}



\maketitle

\section{Introduction}
\label{sec:introduction}

Conversational Recommender Systems (CRSs) have advanced considerably, combining natural language understanding and generation with retrieval, ranking, and adaptive dialogue policies to support multi-turn, context-aware recommendation~\cite{jannach2023evaluating}. Recent pipelines can elicit preferences, justify suggestions, and present relevant alternatives across diverse domains such as e-commerce, education, entertainment, and finance~\cite{gao2021advances, pramod2022conversational}. The emergence of Large Language Models (LLMs) has significantly extended the functional repertoire of these systems; however, the core principles of adaptive recommendation, established well before the current generative AI paradigm, continue to dictate the efficacy of rule-based, statistical, reinforcement-learning, and hybrid approaches~\cite{christakopoulou2016towards, zhao2024recommender}. Accordingly, our focus transcends specific modelling paradigms to address the more fundamental question of how adaptive CRS architectures can be systematically aligned with multi-faceted user experience (UX) requirements.

Despite these technical strides, much CRS research remains \emph{architecture-centric}, primarily emphasising ranking accuracy, intent parsing, or surface-level language fluency \cite{jannach2021survey}. While these dimensions are essential for performance, they often under-represent the fine-grained conversational behaviours and their longitudinal effects on user experience, particularly in systems capable of generating open-ended, non-deterministic content~\cite{deldjoo2404review}. Recent user-centred evaluations have consequently highlighted the necessity of moving beyond coarse outcome measures such as monolithic satisfaction or trust scores toward constructs more specific to the nuances of conversational agency, including perceived understanding, response quality, humanness, and user control~\cite{jin2024crs, mahmud2025evaluating}. This paradigm shift reflects a broader consensus within both the recommender systems and Human-Computer Interaction (HCI) communities: system effectiveness cannot be conceptually or empirically decoupled from experiential quality.

Building on these observations, we propose the \textit{Recommendation-as-Experience} (RAE) framework, which conceptualises recommendation as an unfolding, situated interaction rather than solely a predictive computation. The RAE perspective draws upon foundations in experience-centred HCI~\cite{hassenzahl2010experience}, exploratory search~\cite{marchionini2006exploratory}, and information-seeking theories that foreground the pivotal roles of affect and uncertainty~\cite{kuhlthau2008information}. This framing is strategically aligned with adaptive dialogue management approaches that dynamically adjust strategies in response to latent user goals, emotional cues, and situational contexts~\cite{Berkovsky2016, jameson2007adaptive, mahmud2025understanding}. Within this framework, user experience is operationalised through three complementary \textit{interaction aims}: \textit{educative aims}, which inform and justify to support transparency and decision confidence~\cite{tintarev2015explaining, zhang2020explainable}; \textit{explorative aims}, which inspire and diversify to encourage novelty and serendipity~\cite{white2009exploratory, castells2021novelty, kotkov2016survey}; and \textit{affective aims}, which align emotionally and socially to foster empathy and rapport during the interaction~\cite{okoso2025impact, raamkumar2022empathetic, zhang2024towards}. While each of these aims has been examined in isolation, there remains a critical gap in our understanding of how users evaluate these dimensions \textit{jointly}, or how their relative importance is modulated by domain context, perceived item value, and user-level characteristics. For instance, educative support may dominate in cognitively complex, high-stakes domains such as finance or housing, whereas affective tone may become the primary driver of UX in wellness or mental-health contexts~\cite{sneha2024affective}. Moreover, individual and situational factors including prior CRS experience and specific preferences for dialogue control are known to significantly moderate UX and thus warrant explicit integration into adaptive policy design~\cite{knijnenburg2012explaining, jannach2017user, yang2021designing, pu2011user}.

To investigate these intersections, we conducted a ten-domain study involving $N = 168$ participants, measuring the perceived importance of educative, explorative, and affective aims under varying domain complexities and situational stakes. These predictors correspond to signals that a CRS can feasibly capture at runtime, such as domain metadata, item-level attributes, and persistent user traits. Our analysis provides the empirical grounding necessary to transition from static interfaces to adaptive policies that prioritise specific interaction aims in a context-sensitive manner.

The contributions of this paper are four-fold. First, we provide a multi-domain empirical analysis that quantifies the joint prioritisation of interaction aims across ten distinct CRS contexts. Second, we present empirical evidence that contextual factors including domain-specific cognitive complexity, novelty orientation, and emotional salience systematically moderate user expectations. Third, we identify key user-level influences, establishing prior CRS experience and autonomy preferences as significant modulators while demonstrating that broader demographic effects are largely contingent on specific domain contexts. Finally, we introduce the \textit{RAE adaptation framework}, which provides a formal mapping between structured state signals (domain, value, and user traits) and adaptive interaction policies. By grounding these policies in validated user preferences, the framework advances both the selection of interaction aims (\textit{what} to foreground) and the technical mechanisms of control (\textit{how} to adapt them) for personalised, context-sensitive recommendation. 

\section{Related work}
\label{sec:relatedwork}

Prior research on CRS spans multiple domains, including conversational agents, recommender systems, and human–computer interaction. To situate our contribution, we review four complementary strands: (1) user experience constructs in recommender systems, (2) interaction aims conceptualised as experiential goals, (3) user-level differences and dialogue initiative, and (4) adaptive dialogue policies and orchestration. These research streams reveal both the advances and limitations of existing work, motivating our study and the RAE adaptation framework. Table~\ref{tab:related-work} synthesises representative contributions and highlights research gaps that our study addresses.

\paragraph{User experience constructs in CRS.}
Although CRS research increasingly acknowledges the importance of user experience (UX), most systems remain architecture-centric, optimising ranking accuracy, intent recognition, or language fluency. A growing body of work has expanded evaluation beyond these technical metrics to include constructs such as satisfaction, trust, engagement, and perceived usefulness \cite{manzoor2024chatgpt, jannach2023evaluating, raamkumar2022empathetic}. Instruments such as ResQue \cite{pu2011user} and CRS-Que \cite{jin2024crs} formalise multiple UX dimensions, including informativeness, novelty, transparency, trust, and engagement. These frameworks demonstrate that UX qualities matter for system acceptance and continued use, but they are typically employed as \emph{static outcome criteria} in post-hoc evaluations. 
By contrast, our work treats UX constructs as \emph{dynamic signals} that can be modelled at runtime and incorporated into adaptive state representations. In this perspective, qualities such as transparency or engagement are not only end goals for evaluation but also contextual features that can guide dialogue strategies, retrieval choices, and explanation depth. This reframing moves UX constructs from being merely diagnostic to being operational drivers of adaptive CRS policies.
A recent systematic review of 23 empirical studies across classical and LLM-powered systems found that CRS evaluation remains fragmented, with a significant gap in linking adaptive behaviours to specific UX outcomes. This lack of turn-level affective assessment and the reliance on static post-hoc surveys necessitates a more dynamic approach to experience modelling. The RAE framework addresses this by operationalising these evaluation gaps into active adaptation drivers \cite{mahmud2025evaluating}.

\paragraph{Interaction aims as experiential goals.}
Building on this perspective, recent work has identified specific experiential goals that map to concrete system behaviours. Educative aims emphasise explanations, contextual guidance, and contrastive comparisons, which improve transparency and decision confidence \cite{wang2018explainable, kouki2020generating}. Explorative aims foreground novelty and serendipity, encouraging curiosity-driven discovery and preference formation \cite{kotkov2016survey, chen2021values, marchionini2006exploratory}. Affective aims target rapport, empathy, and sentiment-aware alignment, shown to increase user trust, engagement, and relational satisfaction in domains such as wellness and beauty \cite{okoso2025impact, raamkumar2022empathetic, sharma2021towards}. 
While these aims have each been explored in CRS research, they are typically examined in isolation, either as one-off system features or as independent evaluation dimensions. This fragmentation makes it unclear how aims interact or compete when users face complex decision contexts. Beyond CRS, educational tutors such as ArgueTutor demonstrate that educative, explorative, and affective behaviours can be orchestrated jointly and tailored dynamically to learning goals \cite{wambsganss2021arguetutor}. However, CRS research has yet to establish how users prioritise these aims when they co-occur, or how contextual and individual factors (e.g., domain, task value, user traits) shape their relative salience. 
Our study addresses this gap by quantifying how users jointly value educative, explorative, and affective aims across multiple domains, and by identifying moderators that systematically influence their prioritisation. This contributes empirical grounding for adaptive policies that move beyond single-aim optimisation toward holistic experiential alignment.
Prior empirical work has demonstrated that users equate 'usefulness' and 'effectiveness' in CRS with informational depth and conversational richness, rather than just transactional speed. Furthermore, clustering analyses identified five distinct user profiles with varying preferences for exploratory vs. task-oriented styles. RAE extends this binary style contrast into a tripartite aim structure (Educative, Explorative, Affective) to better capture the nuanced needs of these profiles across diverse domains \cite{mahmud2025understanding}.

\paragraph{User-level differences and dialogue initiative.}
A further gap concerns how user traits and initiative preferences shape experiential priorities in CRS. Demographic moderators such as age and gender have been shown to affect preferences for explanation, transparency, or affective support \cite{huang2024research, rana2021effect, zhang2021effect, okoso2025expressions}. Prior CRS experience also conditions expectations: experienced users often demand greater explanation depth, autonomy, and efficiency, while novice users may require more guidance and reassurance \cite{nourani2020role, komiak2006effects, glikson2020human}. Despite this, few CRS studies stratify samples by experience or test how familiarity systematically modulates the value assigned to different interaction aims \cite{langevin2021heuristic}. Dialogue initiative introduces an additional dimension of variability. Mismatches between a user’s preferred level of initiative (system-led vs. user-led) and the initiative enacted by the CRS reduce trust, usability, and perceived alignment \cite{ma2023initiative, kraus2021modelling, sankaran2020respecting}. Reviews of dialogue management further identify initiative handling as a persistent challenge for adaptive systems \cite{brabra2021dialogue}. 
These findings indicate that adaptive CRS must not only respond to contextual signals such as domain and task value, but also incorporate user-level traits and initiative preferences into their interaction policies. Our study contributes by testing how demographic factors, prior CRS experience, and autonomy preferences influence the prioritisation of educative, explorative, and affective aims across domains.

\paragraph{Adaptive dialogue policies and orchestration.}
Adaptive dialogue policies provide the computational machinery that maps contextual and user signals into behaviour selection. Classical approaches employed POMDPs and policy gradient methods to optimise dialogue under uncertainty \cite{young2013pomdp, sun2018conversational}, and CRS-specific pipelines interleaved preference elicitation with item retrieval \cite{christakopoulou2016towards}. Reinforcement learning has been used to optimise long-term engagement, efficiency, and reward-driven outcomes \cite{yang2023hierarchical, gao2021advances}, with recent advances introducing multi-round feedback adaptation \cite{xu2021adapting}, graph-based policy unification \cite{chen2021knowledge}, and knowledge-grounded CRS with policy learning \cite{deng2021unified}. 
More recently, LLM-assisted orchestration integrates retrieval, reasoning, and controllable generation for adaptive recommendation \cite{lin2025can, xu2024leveraging}, with surveys emphasising the importance of explicit orchestration layers and state representations \cite{wu2024survey}. These approaches illustrate the growing sophistication of adaptive dialogue policies, but most continue to optimise for efficiency or accuracy, rarely encoding experiential aims as first-class state variables. 
Our work extends this trajectory by formalising a \emph{state--policy--behaviour} mapping in which educative, explorative, and affective aims are explicitly represented as adaptive features. This shift reframes dialogue management from a performance-centric optimisation task to one that also foregrounds experiential quality, bridging system-level orchestration with validated user preferences.

\begin{table}[h]
\centering
\begin{threeparttable}
\caption{Taxonomy of research streams in Conversational Recommender Systems (CRS) and identified gaps. 
The present study addresses these limitations by modelling the joint prioritisation of interactional aims and formalising a state--policy--behaviour framework for adaptive, experience-centric CRS design.}
\label{tab:related-work}

\footnotesize 
\begin{tabularx}{\textwidth}{@{} p{3.8cm} X @{}}
\toprule
\textbf{Research Stream} & \textbf{Typical Emphasis and Identified Research Gap} \\
\midrule
\textbf{User Experience Constructs} & Focus on measurement instruments (e.g., ResQue \cite{pu2011user}, CRS-Que \cite{jin2024crs}) and outcomes such as trust, engagement, and perceived usefulness \cite{manzoor2024chatgpt, jannach2023evaluating}. \textit{Gap: Constructs are typically treated as static outcomes rather than dynamic state features for real-time adaptation.} \\
\addlinespace
\textbf{Interaction Aims} & Educative (explanations, trade-off comparisons) \cite{wang2018explainable, kouki2020generating}; Explorative (diversity, novelty) \cite{kotkov2016survey, chen2021values}; and Affective (tone, empathy) support \cite{okoso2025impact, raamkumar2022empathetic}. \textit{Gap: The joint prioritisation and trade-offs between these three aims across different decision domains remains largely unstudied.} \\
\addlinespace
\textbf{User Traits and Initiative} & Impact of demographics (age, gender) \cite{huang2024research, rana2021effect} and prior CRS experience \cite{nourani2020role, glikson2020human} on dialogue initiative preferences \cite{ma2023initiative, sankaran2020respecting}. \textit{Gap: Insufficient evidence regarding how individual user traits and initiative preferences should modulate specific interactional aims.} \\
\addlinespace
\textbf{Adaptive Policies} & Classical RL (POMDP) \cite{young2013pomdp}, hierarchical efficiency \cite{yang2023hierarchical}, and LLM orchestration for controllable generation \cite{lin2025can, he2023large}. \textit{Gap: Current models prioritise transactional efficiency or retrieval accuracy but rarely encode multi-dimensional experiential aims into state--policy mappings.} \\
\bottomrule
\end{tabularx}
\end{threeparttable}
\end{table}

\section{Method}
\label{sec:methods}

We conducted a cross-sectional online survey to examine how users prioritise \emph{educative}, \emph{explorative}, and \emph{affective} interaction aims across multiple CRS domains, and how these priorities vary with perceived item value, user characteristics, and dialogue-control preferences. The measured variables align with features that can be operationalised in a CRS \emph{state representation}, enabling rule-based or learned policies in LLM-powered and other adaptive CRS architectures to adjust behaviours in real time.

\subsection{Study design}
\label{sec:study-design}

The study employed a vignette-based survey design to elicit user priorities across ten diverse CRS application domains. This approach was selected over a functional system deployment to ensure comparative generalisability across distinct cognitive, emotional, and social contexts while avoiding the technical confounds associated with specific system architectures. Our design addresses three primary research questions operationalised through six hypotheses as described in Section~\ref{sec:research-hypotheses}.

\textit{Domain Selection and Categorisation.} 
Ten domains were selected based on their prevalence in current CRS deployments and their representation in recent state-of-the-art reviews~\cite{pramod2022conversational, jannach2021survey, jin2024crs}. To ensure a broad coverage of the recommendation landscape, we categorised these domains into five functional groups:
\begin{itemize}[leftmargin=1.5em, nosep, label={--}]
  \item High-stakes and Complex: Finance, Housing, and Education.
  \item Hedonic and Leisure: Travel, Apparel, and Entertainment.
  \item Affect-rich and Identity-based: Wellness and Beauty.
  \item Functional and Technical: Technology (consumer electronics).
  \item Service and Hospitality: Dining.
\end{itemize}

\textit{Sample Size and Power Analysis.} 
An \textit{a priori} power analysis was conducted using G*Power 3.1~\cite{faul2007g}. To detect a medium effect size ($f = 0.25$) in an omnibus test with a significance level of $\alpha = .05$ and power of $1 - \beta = .80$, a minimum sample of $N = 159$ was required. While our primary analyses utilise non-parametric and Bayesian ordinal models, we adopted this target as a conservative baseline for ensuring sufficient sensitivity to detect cross-domain differences.

\textit{Methodological Rationale.} 
By adopting a vignette-based elicitation approach, the study identifies framework-level expectations that are architecture-agnostic. This design allows for the isolation of the "RAE state variables" (domain, user traits, and item value) without interference from dialogue-system latency, natural language understanding errors, or specific interface aesthetics, thereby grounding the framework in high-level user requirements.

\subsection{Research hypotheses}
\label{sec:research-hypotheses}

Based on the proposed RAE framework, we formulated six hypotheses linking contextual and individual factors to the perceived importance of interactional aims. These hypotheses focus on signals that a CRS can dynamically capture as state variables to drive adaptive policies.

\paragraph{RQ1: Domain effects on interactional aims}
The first set of hypotheses examines how the inherent characteristics of a decision domain shape baseline user expectations:
\begin{description}[leftmargin=1.5em, nosep, font=\normalfont\itshape]
  \item[H1:] Perceived importance of the \textit{educative} aim is significantly moderated by domain-specific task complexity.
  \item[H2:] Perceived importance of the \textit{explorative} aim is significantly moderated by domain-specific novelty orientation.
  \item[H3:] Perceived importance of the \textit{affective} aim is significantly moderated by domain-specific emotional relevance.
\end{description}

\paragraph{RQ2: Moderation by item value and user traits}
The second set of hypotheses investigates how situational stakes and individual differences modulate the baseline domain effects:
\begin{description}[leftmargin=1.5em, nosep, font=\normalfont\itshape]
  \item[H4:] Situational \textit{item value} significantly impacts user prioritisation of educative, explorative, and affective aims, with higher stakes increasing demand across all dimensions.
  \item[H5:] User characteristics—specifically \textit{age}, \textit{gender}, and \textit{CRS experience}—significantly moderate the perceived importance of interactional aims.
\end{description}

\paragraph{RQ3: Dialogue-control preferences}
The final set of hypotheses explores the interactional agency users desire during these exchanges:
\begin{description}[leftmargin=1.5em, nosep, font=\normalfont\itshape]
  \item[H6a:] Users exhibit a significant preference for \textit{user-initiated control} during both educative and explorative interaction phases (rated significantly above the neutral midpoint).
  \item[H6b:] Demographic traits and prior experience are significantly associated with the degree of control-preference for these interactional aims.
\end{description}

\subsection{Survey instrument and measures}
\label{sec:survey-instrument}

The research instrument was developed as a structured digital questionnaire designed to capture multi-dimensional user expectations across diverse recommendation contexts. The instrument was grounded in established user-centric evaluation frameworks for recommender systems~\cite{pu2011user, knijnenburg2012explaining} and specific metrics for conversational interaction~\cite{jin2024crs, mahmud2025evaluating}, ensuring compatibility with adaptive design parameters and multi-dimensional UX requirements~\cite{zhang2020explainable, chen2021values}.

\textit{Elicitation of interactional aims.} 
Central to the instrument was the elicitation of latent user preferences across three interactional aims: \textit{Educative} (the requirement for transparent rationale), \textit{Explorative} (the demand for discovery and serendipity), and \textit{Affective} (the expectation for emotional and social alignment). These constructs were operationalised as single-item measures to ensure high completion quality and mitigate respondent fatigue across the intensive 10-domain block. For example, the educative aim was operationalised through the item: \textit{'How important is it that the system explains why these options were chosen?'} (measured on a 5-point Likert scale). To neutralise potential order effects, both the sequence of domains and the presentation of interactional aims were fully randomised for each participant.

\textit{Situational item-value manipulation.} 
To examine the within-subjects effect of situational stakes (H4), the instrument featured contrasted vignettes. Participants were presented with situated scenarios that juxtaposed low-stakes contexts (e.g., a casual \textdollar 10 purchase) against high-stakes conditions (e.g., a significant \textdollar 1,000 investment). This within-subjects design ensured that 'perceived value' was anchored in realistic, domain-specific consumer behaviours rather than abstract estimation, allowing for the isolation of value-sensitivity as a dynamic modulator.

\textit{User traits and autonomy preferences.} 
The final component of the instrument captured demographic covariates and dialogue-control expectations. We recorded age group, gender identity, and prior CRS experience, the latter measured on a 5-point frequency scale. Furthermore, dialogue-control preferences for \textit{educative} and \textit{explorative} aims were elicited using a 5-point bipolar scale (ranging from 1 = \textit{fully system-initiated} to 5 = \textit{fully user-initiated}). These items were framed within a familiar apparel e-commerce context to provide a stable reference point for assessing autonomy. Consistent with our theoretical framework, no autonomy item was collected for the affective aim, as affective support is conceptually defined as a system-led response to implicit user cues. 
The complete instrument underwent a pre-test phase ($n = 8$) to refine linguistic clarity and cognitive load. Items were strategically adapted from \textit{CRS-Que}~\cite{jin2024crs} and validated UX metrics~\cite{zhang2020explainable, chen2021values, raamkumar2022empathetic} to maintain high construct validity.

\paragraph{Participant recruitment and data refinement}

We recruited a total of $N = 191$ participants through a multi-channel strategy including Prolific ($n = 149$), Survey Swap ($n = 33$), and targeted social media recruitment ($n = 9$). Eligibility was restricted to individuals aged 18 or older with verified English proficiency. Prolific participants were further screened to include only those with a high historical performance record ($\geq 95\%$ approval rating and $\geq 100$ completed studies). The survey was hosted on the Qualtrics platform, with an average completion time of \SI{9}{\minute}; Prolific participants were compensated \textdollar 2.40, a rate consistent with institutional and platform ethical guidelines.
To ensure data integrity, we applied rigorous cleansing procedures to identify Insufficient Effort Responding (IER)~\cite{meade2012identifying}. A total of 23 cases were excluded based on the following pre-defined criteria: failing the embedded attention checks (9 cases), exhibiting 'straight-lining' behaviour with a standard deviation across Likert blocks of $SD < 0.2$ (7 cases), or presenting implausibly rapid completion times of less than \SI{3}{\minute} (7 cases). The final analytic sample comprised $n = 168$ valid respondents. This sample size retained sufficient statistical power for the planned frequentist and Bayesian analyses and provided a diverse distribution of age, gender, and CRS experience necessary for robust moderator analysis.

\subsection{Data analysis}
\label{sec:data-analysis}

Analyses were conducted in Python~3.11 (\texttt{pandas 2.2}, \texttt{scipy 1.12}, \texttt{statsmodels 0.15}); Bayesian models were fitted in \texttt{Stan 2.33} via \texttt{cmdstanpy}. Ordinal responses (1–5) were analysed using non-parametric tests or ordinal regression as appropriate. All tests were two-sided with $\alpha = .05$. Missing data were handled listwise \emph{per analysis} (i.e., pairwise inclusion), and sample sizes ($n$) are reported with each result. For H6 regressions, the effective model sample was $N=164$ after listwise inclusion of covariates.

\paragraph{Domain effects (H1–H3).}
Kruskal–Wallis tests \cite{kruskal1952use}were followed, where appropriate, by Bonferroni-adjusted Mann–\allowbreak Whitney $U$ comparisons. We also fitted hierarchical Bayesian cumulative-logit models \cite{gelman2013philosophy}with domain as a fixed factor and participant-level random intercepts to estimate domain-level uncertainty. Pairwise effect sizes are reported as $r = |z|/\sqrt{n}$.

\paragraph{Item value (H4).}
Within-subject differences between low- and high-value frames were tested with Wilcoxon signed-rank tests \cite{wilcoxon1945individual}. For transparency, we report Wilcoxon $W$, the non-tied sample size $n'$ (ratings equal to the midpoint are dropped by the test when applicable), $p$-values, and effect sizes ($r=|z|/\sqrt{n'}$). Rank-biserial and common-language effect size (CLES) are provided.

\paragraph{Demographics (H5).}
Zero-order associations were assessed with Spearman rank correlations; the Benjamini–Hoc\-hberg procedure \cite{benjamini1995controlling} controlled the false discovery rate at $q=.05$ \emph{per predictor family} (30 correlations per demographic predictor). Bayesian cumulative-logit models with random domain intercepts provided pooled estimates across domains (weakly informative, zero-centred priors; diagnostics reported in the supplement).

\paragraph{Autonomy preferences (H6a–H6b).}
One-sample Wilcox\-on signed-rank tests against the neutral midpoint ($=3$) assessed preferences for user initiation in educative and explorative aims; ties at the midpoint were dropped by the test ($n'$ reported). Separate cumulative-logit models \cite{hosmer2013applied} examined demographic predictors, with Gender coded Male$=1$, Female$=0$, Age treated as categorical (reference $=$ 18–24), and CRS experience modelled as ordinal (1–5). Proportional-odds assumptions and model fit were assessed via surrogate diagnostics and posterior predictive checks.

\paragraph{Inference criteria and diagnostics.}
Frequentist significance was defined as $p \le .05$. For Bayesian models, effects were deemed credible when 95\% posterior intervals excluded zero. Convergence and adequacy were assessed using $\hat{R}$, effective sample sizes, divergence checks, and posterior predictive fit.
Given the use of single-item measures, we prioritised Bayesian credible intervals (95\% HDI) over internal consistency metrics (e.g., Cronbach's $\alpha$). The Bayesian approach provides a robust estimate of parameter uncertainty and effect stability, which serves as a statistically rigorous proxy for reliability in multi-domain elicitation studies.

\subsection{Ethics}
\label{sec:ethics}

The study was approved by the authors’ institutional Human Research Ethics Committee. All participants provided informed consent. No personally identifiable data were collected; data were anonymised at source and stored in compliance with GDPR and institutional data-governance policies.

\section{Results}
\label{sec:results}

This section presents the empirical findings from our vignette-based elicitation study ($N=168$). We begin by detailing the participant profile, followed by a systematic evaluation of our six hypotheses (H1--H6) using a combination of frequentist non-parametric tests and Bayesian hierarchical models. Collectively, the data validate the core premise of the RAE framework: interactional aims are significantly modulated by the decision domain, situational item value, and prior system experience. A consolidated summary of all hypothesis outcomes is provided in Table~\ref{tab:hypotheses-summary} at the end of this section.

\subsection{Participant demographics}
Table~\ref{tab:demographics} provides a breakdown of the study population. The sample primarily represents younger to middle-aged adults, with \SI{73.2}{\percent} of participants aged 18--34 years. In terms of gender identity, the sample comprised \SI{57.1}{\percent} female, \SI{40.5}{\percent} male, and \SI{1.8}{\percent} non-binary or other. Notably, the cohort was nearly evenly split regarding system familiarity: \SI{52.4}{\percent} reported at least some prior experience with Conversational Recommender Systems (CRS), while \SI{47.6}{\percent} were novices. Within the RAE framework, these attributes serve as persistent user-state features that inform the baseline dialogue policy.

\begin{table}[h]
\centering
\caption{Demographic profile and baseline system experience of the study population ($N = 168$). The sample represents a broad distribution of young and middle-aged adults, with a balanced split between users with and without prior Conversational Recommender System (CRS) experience.}
\label{tab:demographics}
\begin{threeparttable}
\footnotesize
\setlength{\tabcolsep}{23pt} 
\begin{tabular}{ll r S[table-format=2.1]}
\toprule
\textbf{Characteristic} & \textbf{Category} & \textbf{$n$} & {\textbf{\%}} \\
\midrule
Age Group       & 18--24 years & 58 & 34.5 \\
                & 25--34 years & 65 & 38.7 \\
                & 35--44 years & 31 & 18.5 \\
                & 45--54 years & 8  & 4.8  \\
                & 55--64 years & 6  & 3.6  \\
                & 65+ years    & 0  & 0.0  \\
\addlinespace[4pt]
Gender Identity & Female       & 96 & 57.1 \\
                & Male         & 68 & 40.5 \\
                & Other (incl.\ non-binary) & 3 & 1.8 \\
                & Prefer not to say & 1 & 0.6 \\
\addlinespace[4pt]
CRS Experience  & None         & 80 & 47.6 \\
                & Some / Frequent & 88 & 52.4 \\
\bottomrule
\end{tabular}
\begin{tablenotes}
    \scriptsize
    \item \textit{Note:} CRS Experience was self-reported based on prior interaction with AI-driven travel or shopping assistants. 
    \item Percentages are calculated row-wise and may not sum to 100\% due to rounding. 
    CRS Experience was self-reported based on prior interaction with AI-driven travel or shopping assistants.
\end{tablenotes}
\end{threeparttable}
\end{table}

\subsection{Domain effects on interaction preferences (H1--H3)}
All three interaction aims varied significantly across domains, as shown in Table~\ref{tab:interaction-aims-by-domain}, with Education, Travel, and Wellness emerging as the top-ranked domains for educative, explorative, and affective aims, respectively. Furthermore, we conducted Kruskal--Wallis H tests to evaluate domain-level differences for all three interaction aims. All three yielded significant omnibus effects: Educative ($H(9) = 93.15$, $p < .001$), Explorative ($H(9) = 33.34$, $p < .001$), and Affective ($H(9) = 25.10$, $p = .003$), indicating that users’ prioritisation of each aim differed significantly by domain. Table~\ref{tab:kruskal-wallis-results} shows the test statistics and ranks the top and bottom three domains for each aim by mean rank. Follow-up pairwise comparisons using Bonferroni-adjusted Mann--Whitney U tests identified several domain pairs with meaningful differences ($r \geq 0.20$). As summarised in Table~\ref{tab:pairwise-differences}, the largest effect emerged between Education and Dining for the Educative aim ($r = .36$), supporting the hypothesis that educative features are more valued in complex, high-stakes domains. For the Explorative aim, Travel was rated significantly higher than Wellness ($r = .25$), consistent with novelty-seeking preferences in exploratory domains. For the Affective aim, Wellness was rated higher than Tech ($r = .21$), reflecting heightened affective expectations in emotionally salient domains.

\begin{table}[h]
\centering
\caption{Descriptive statistics for perceived interactional aims across ten decision domains ($N=168$). Values represent the Median (Med.), Mode, and Interquartile Range (IQR). 
Notably, high-stakes domains such as Travel, Finance, and Housing consistently elicit higher central tendency scores for Educative support, while hedonic domains like Apparel and Entertainment show a stronger median preference for Explorative interaction.}
\label{tab:interaction-aims-by-domain}
\begin{threeparttable}
\footnotesize
\setlength{\tabcolsep}{0pt}
\begin{tabular*}{\textwidth}{@{\extracolsep{\fill}} l ccc @{\hspace{15pt}} ccc @{\hspace{15pt}} ccc}
\toprule
\textbf{Domain} & \multicolumn{3}{c}{\textbf{Educative}} & \multicolumn{3}{c}{\textbf{Explorative}} & \multicolumn{3}{c}{\textbf{Affective}} \\
\cmidrule(lr){2-4} \cmidrule(lr){5-7} \cmidrule(lr){8-10}
 & Med. & Mode & IQR & Med. & Mode & IQR & Med. & Mode & IQR \\
\midrule
Apparel       & 3.00 & 4 & 3--4 & 4.00 & 4 & 3--4 & 3.00 & 3 & 2--4 \\
Beauty        & 4.00 & 4 & 3--4 & 3.00 & 4 & 3--4 & 3.00 & 4 & 2--4 \\
Entertainment & 3.00 & 4 & 3--4 & 4.00 & 4 & 3--4 & 3.00 & 3 & 3--4 \\
Tech          & 4.00 & 4 & 3--5 & 4.00 & 4 & 3--5 & 3.00 & 3 & 2--4 \\
Dining        & 3.00 & 4 & 3--4 & 4.00 & 4 & 3--4 & 3.00 & 3 & 3--4 \\
Wellness      & 4.00 & 4 & 3--4 & 3.00 & 3 & 3--4 & 4.00 & 4 & 3--4 \\
Travel        & 4.00 & 5 & 3--5 & 4.00 & 4 & 3--5 & 4.00 & 5 & 3--5 \\
Education     & 4.00 & 4 & 3--5 & 4.00 & 4 & 3--5 & 3.00 & 4 & 2--4 \\
Finance       & 4.00 & 4 & 3--5 & 4.00 & 4 & 3--4 & 3.00 & 4 & 2--4 \\
Housing       & 4.00 & 4 & 3--4 & 4.00 & 4 & 3--4 & 3.00 & 3 & 2--4 \\
\bottomrule
\end{tabular*}

\begin{tablenotes}
    \scriptsize
    \item \textit{Note:} Ratings measured on a 5-point Likert scale (1 = Low priority, 5 = High priority). IQR reported as 25th--75th percentiles.
\end{tablenotes}
\end{threeparttable}
\end{table}

\begin{table}[h]
\centering
\caption{Kruskal--Wallis $H$-test results and domain-level mean ranks for interactional aims ($N=168$ users across 10 domains). Significant $H$-statistics across all three aims ($p \le .003$) confirm that user prioritisation of Educative, Explorative, and Affective support is context-dependent. Mean ranks highlight the divergence between high-stakes/functional domains (e.g., Education, Tech) and hedonic/wellness domains.} 
\label{tab:kruskal-wallis-results}
\begin{threeparttable}
\footnotesize
\setlength{\tabcolsep}{15pt}
\begin{tabular}{l ccc p{3.2cm} p{3.2cm}}
\toprule
\textbf{Interaction} & \textbf{$H$} & \textbf{$df$} & \textbf{$p$} & \textbf{Top 3 Domains} & \textbf{Bottom 3 Domains} \\
\textbf{Aim} & & & & \textbf{(Mean Rank)} & \textbf{(Mean Rank)} \\
\midrule
Educative   & 93.15 & 9 & $<.001$ & Education (1014.9) \par Tech (981.8) \par Travel (960.9) & Dining (681.0) \par Apparel (730.4) \par Entert. (732.5) \\
\addlinespace[8pt]
Explorative & 33.34 & 9 & $<.001$ & Travel (948.9) \par Education (916.4) \par Tech (898.3) & Wellness (730.4) \par Beauty (770.5) \par Dining (792.9) \\
\addlinespace[8pt]
Affective   & 25.10 & 9 & .003    & Wellness (922.4) \par Travel (918.0) \par Entert. (885.2) & Tech (720.6) \par Housing (779.5) \par Finance (822.4) \\
\bottomrule
\end{tabular}
\begin{tablenotes}
    \scriptsize
    \item \textit{Note:} $H$ follows a $\chi^2$ distribution with 9 degrees of freedom ($df$). Mean ranks represent the relative prioritisation after pooling all user responses ($n=1680$ observations). 
    Higher ranks indicate a stronger demand for that specific interactional quality.
\end{tablenotes}
\end{threeparttable}
\end{table}

\begin{table}[h]
\centering
\caption{Summary of significant pairwise differences from Bonferroni-adjusted Mann--Whitney $U$ tests. 
Only pairs with a notable effect size ($r \ge 0.20$) are reported. The results indicate that Educative support requirements exhibit the strongest differentiation between domains (e.g., Education vs. Dining), while Affective requirements are uniquely prioritised in Wellness and Travel contexts.}
\label{tab:pairwise-differences}
\begin{threeparttable}
\footnotesize
\setlength{\tabcolsep}{18pt}
\begin{tabular}{l p{6cm} c}
\toprule
\textbf{Interaction Aim} & \textbf{Significant Domain Pairs (Directional)} & \textbf{Effect Size ($r$)} \\
\midrule
Educative   & Education $>$ Dining; Education $>$ Apparel; \par Tech $>$ Dining; Tech $>$ Apparel; Travel $>$ Dining & 0.25--0.36 \\
\addlinespace[8pt]
Explorative & Travel $>$ Wellness; Education $>$ Wellness & 0.23--0.25 \\
\addlinespace[8pt]
Affective   & Tech $<$ Wellness; Tech $<$ Travel & 0.21--0.22 \\
\bottomrule
\end{tabular}
\begin{tablenotes}
    \scriptsize
    \item \textit{Note:} $p$-values were Bonferroni-corrected. Effect sizes $r$ are interpreted via Cohen's guidelines: $\ge 0.10$ (small), $\ge 0.30$ (medium). A $>$ symbol indicates the first domain in the pair was rated significantly higher for that specific interaction aim.
\end{tablenotes}
\end{threeparttable}
\end{table}

We further validated these findings using Bayesian hierarchical ordinal regression models, which incorporated domain-level random intercepts and individual-level fixed effects. Posterior estimates confirmed the frequentist results. For example, Education domain had the highest estimated intercept for educative interaction ratings ($\beta_{\text{Education}} = 1.25$, 94\% HDI [0.47, 1.96]), while Travel had the highest for explorative interaction ratings ($\beta_{\text{Travel}} = 1.30$, 94\% HDI [0.52, 1.99]), and Wellness for affective interaction ratings ($\beta_{\text{Wellness}} = 1.12$, 94\% HDI [0.38, 1.83]).
These results provide robust support for H1, H2, and H3. User preferences for educative, explorative, and affective interaction aims are not uniformly distributed but vary systematically across application domains. While our analysis does not directly measure latent factors such as domain complexity, novelty orientation, or emotional salience, the observed patterns are consistent with these theoretical characteristics. Although interaction demands may differ across specific items within a domain, the present results indicate domain-level user expectations that are robust enough to inform adaptive CRS design. These domain effects can serve as priors for domain-conditioned policies, enabling systems to initialise aim weightings and explanation styles according to task context.

\subsection{Impact of item value on interaction aims (H4)}
\label{subsec:item-value}
\begin{figure}
  \centering
  \includegraphics[scale=0.65]{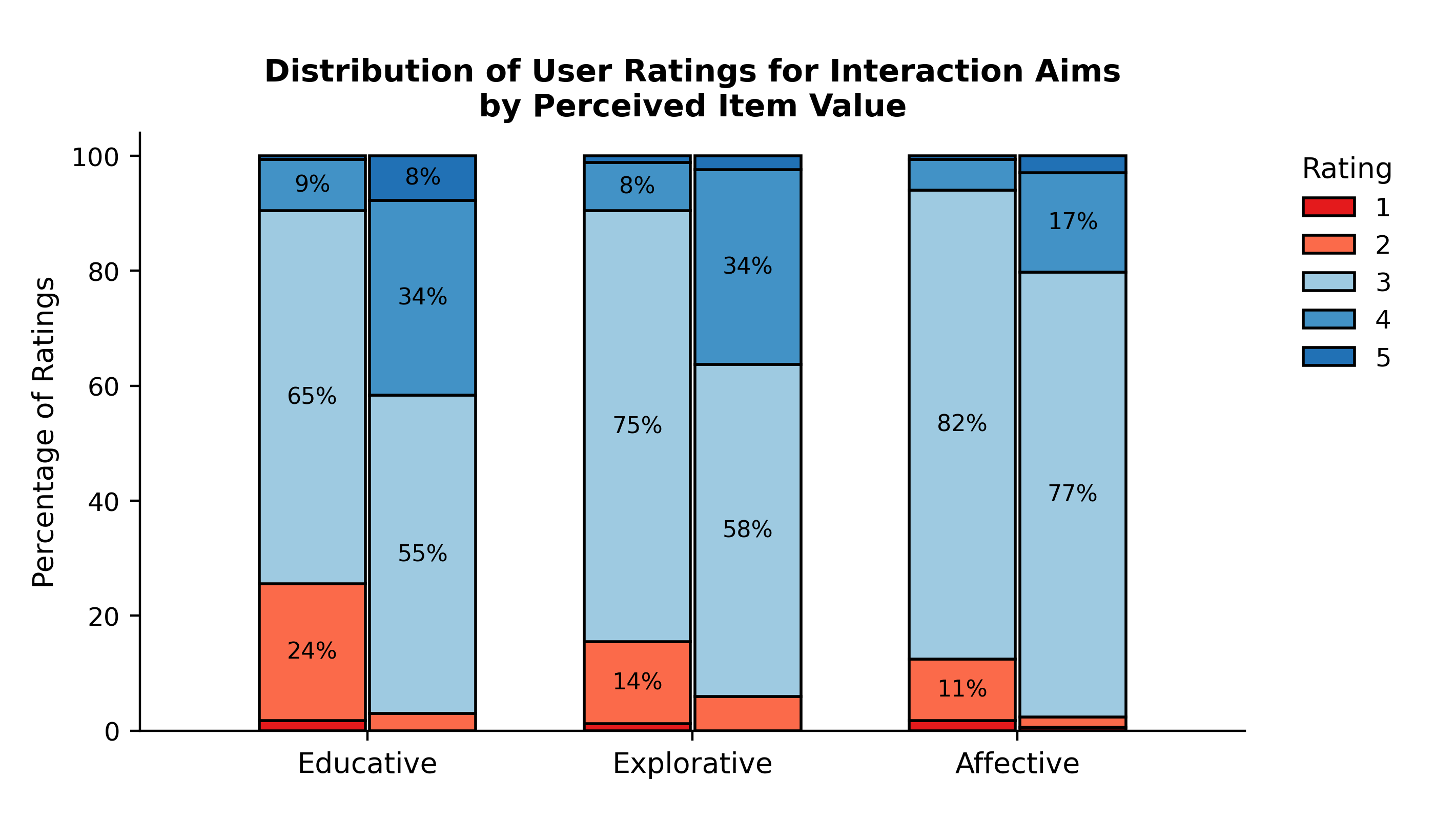}
  \caption{Stacked-bar distribution (\%) of user ratings for educative, explorative, and affective interaction aims, comparing low-value and high-value item scenarios. Bars on the \textit{left} represent low-value items; bars on the \textit{right} represent high-value items. Colour gradients from red (rating = 1) to dark blue (rating = 5) indicate increasing perceived importance.}
  \label{fig:item-value-stacked}
\end{figure}

To evaluate H4, participants rated the perceived importance of educative, explorative, and affective aims across two distinct situational contexts: a low-value and a high-value item scenario. This yielded $n = 168$ paired observations per interactional aim. Figure~\ref{fig:item-value-stacked} illustrates the distribution of these ratings. 
Across all three aims, the transition to a high-value scenario produced a pronounced upward shift in user expectations. The proportion of low-importance ratings (1--2) decreased substantially, while high-importance ratings (4--5) saw a corresponding surge. For the educative aim, high-importance ratings rose from \SI{9}{\percent} to \SI{42}{\percent} (a 33~percentage point increase), while the affective aim saw a shift from \SI{1}{\percent} to \SI{18}{\percent}.
Two-sided Wilcoxon signed-rank tests confirmed that these increases were statistically significant across all dimensions: 
educative ($W = 368.5$, $z = -10.66$, $p < .001$, $r = .82$), 
explorative ($W = 473.5$, $z = -10.49$, $p < .001$, $r = .81$), and 
affective ($W = 68.5$, $z = -11.13$, $p < .001$, $r = .86$). 
All observed effects were categorised as large according to conventional benchmarks. While median scores remained at 3.0, this measure of central tendency masked the substantial distributional shifts evident in both the stacked-bar visualisation and the large effect sizes.
These findings provide robust support for H4, suggesting that perceived decision stakes significantly elevate the importance users place on all three interactional aims. Within the RAE framework, item value functions as a high-priority \textit{state variable} that should trigger real-time policy adjustments. Specifically, under high-value conditions, the dialogue policy should scale behavioral intensity across three fronts: \textit{educative behaviours} should transition toward deeper, more contextualised explanations and comparative justifications; \textit{explorative behaviours} should expand to include broader, more diverse retrieval sets; and \textit{affective behaviours} should adopt a more empathetic, alignment-oriented tone. This mapping ensures that system responses, whether implemented via prompt conditioning or controllable NLG, are scaled proportionally to the situational stakes inferred from task metadata.

\FloatBarrier

\begin{table}[ht]
\centering
\caption{Wilcoxon signed-rank test results comparing user prioritisation of interactional aims between low-value and high-value item conditions ($n = 168$ pairs). 
The extremely large effect sizes across all dimensions ($r > .80$) indicate that the perceived value of the recommendation item is a primary driver of user demand for Educative, Explorative, and Affective support, justifying the adaptive state-logic of the RAE framework.}
\label{tab:h4-wilcoxon}
\begin{threeparttable}
\footnotesize 
\setlength{\tabcolsep}{18pt} 
\begin{tabular}{l cccc}
\toprule
\textbf{Interaction Aim} & \textbf{$W$} & \textbf{$z$} & \textbf{$p$} & \textbf{$r$} \\
\midrule
Educative   & 368.5 & $-10.66$ & $<.001$ & $.822$ \\
Explorative & 473.5 & $-10.49$ & $<.001$ & $.809$ \\
Affective   & 68.5  & $-11.13$ & $<.001$ & $.859$ \\
\bottomrule
\end{tabular}
\begin{tablenotes}
    \scriptsize
    \item \textit{Note:} All tests are paired and two-tailed. Effect size $r$ is computed as $|z| / \sqrt{n}$. 
    According to Cohen’s guidelines, $r > .5$ represents a large effect.
\end{tablenotes}
\end{threeparttable}
\end{table}

\subsection{Moderation by user characteristics (H5)}
\label{sec:demographics}

Hypothesis 5 posited that individual differences, specifically \textit{age}, \textit{gender}, and prior \textit{CRS experience}, would predict user prioritisation of interactional aims. To test this, we first conducted an exploratory domain-level correlation screen with Benjamini--Hochberg (FDR) correction, followed by confirmatory Bayesian hierarchical models to estimate global effects across domains.

\paragraph{Domain-level associations (Spearman).}
For this analysis, gender was effect-coded (female = $+1$, male = $-1$) and CRS experience was treated as an ordinal 5-point scale. Table~\ref{tab:demographic-correlations} reports the resulting Spearman’s $\rho$ coefficients and FDR-corrected $p$-values.
\textit{CRS experience} emerged as the most robust predictor, with 13 of 30 domain-interaction combinations reaching significance after correction. This included educative importance in \textit{Dining} and \textit{Wellness}; explorative importance across almost all domains except \textit{Tech} and \textit{Entertainment}; and affective importance in \textit{Apparel}, \textit{Beauty}, \textit{Finance}, and \textit{Tech}. In contrast, \textit{gender} effects were highly localized, appearing only for explorative aims in \textit{Finance} and affective aims in \textit{Housing}. Notably, \textit{age} yielded no significant associations across any domain or aim after multiple-comparison correction.

\begin{table}[ht]
\centering
\caption{Correlation Analysis of Demographic Covariates and Interaction Aims. The table presents Spearman's rank correlation coefficients ($\rho$) and Benjamini--Hochberg (BH) adjusted $p$-values. Notably, CRS Experience shows a consistent positive association, whereas Age and Gender exhibit minimal significant effects.}
\label{tab:demographic-correlations}
\begin{threeparttable}
\footnotesize 
\setlength{\tabcolsep}{12pt} 
\begin{tabular}{ll r@{\hspace{4pt}}l r@{\hspace{4pt}}l r@{\hspace{4pt}}l}
\toprule
\textbf{Domain} & \textbf{Interaction} & \multicolumn{2}{c}{\textbf{Age}} & \multicolumn{2}{c}{\textbf{Gender}} & \multicolumn{2}{c}{\textbf{CRS Exp.}} \\
& \textbf{Aim} & \multicolumn{2}{c}{$\rho$ ($p$)} & \multicolumn{2}{c}{$\rho$ ($p$)} & \multicolumn{2}{c}{$\rho$ ($p$)} \\
\midrule
Apparel    & Explorative & $-0.04$ & (.99) & $-0.01$ & (.93) & $0.27^{**}$ & (.00) \\
Education  & Educative   & $-0.15$ & (.78) & $-0.00$ & (.99) & $0.18^{*}$  & (.05) \\
Education  & Explorative & $-0.07$ & (.99) & $0.14$  & (.26) & $0.19^{*}$  & (.04) \\
Entertain. & Affective   & $-0.01$ & (.99) & $-0.02$ & (.92) & $0.24^{*}$  & (.01) \\
Finance    & Affective   & $0.02$  & (.99) & $0.14$  & (.26) & $0.21^{*}$  & (.03) \\
Finance    & Educative   & $0.00$  & (.99) & $0.04$  & (.77) & $0.19^{*}$  & (.04) \\
Finance    & Explorative & $0.10$  & (.99) & $0.24^{*}$ & (.04) & $0.20^{*}$ & (.04) \\
Housing    & Affective   & $-0.00$ & (.99) & $0.24^{*}$ & (.04) & $0.16$      & (.07) \\
Housing    & Explorative & $-0.06$ & (.99) & $0.19$  & (.15) & $0.18^{*}$  & (.05) \\
Tech       & Affective   & $0.04$  & (.99) & $0.07$  & (.70) & $0.32^{**}$ & (.00) \\
Tech       & Explorative & $-0.02$ & (.99) & $0.04$  & (.77) & $0.20^{*}$  & (.04) \\
Wellness   & Affective   & $-0.16$ & (.78) & $0.14$  & (.26) & $0.23^{*}$  & (.01) \\
Wellness   & Educative   & $-0.00$ & (.99) & $0.04$  & (.77) & $0.24^{*}$  & (.01) \\
Wellness   & Explorative & $-0.02$ & (.99) & $0.12$  & (.33) & $0.33^{**}$ & (.00) \\
\bottomrule
\end{tabular}
\begin{tablenotes}
    \scriptsize
    \item \textit{Note:} Gender effect-coded ($+1$ Female, $-1$ Male). CRS Exp. measured 1--5. Significance: $^{*} p < .05, ^{**} p < .01$ (BH-adjusted). Only rows with at least one significant result shown.
\end{tablenotes}
\end{threeparttable}
\end{table}

\paragraph{Bayesian hierarchical ordinal models.}
\label{sec:results-bayesian-models}

To confirm these trends, we fitted cumulative-logit mixed models using \texttt{Stan} (4 chains, \num{2000} post-warmup draws). These models incorporated domain-level random intercepts ($\alpha_j \sim \mathcal{N}(0,\sigma_{\alpha})$) and weakly informative $\mathcal{N}(0,1)$ priors. Convergence was excellent, with $\hat{R} < 1.01$ and effective sample sizes (ESS) exceeding \num{1500} for all parameters. Posterior predictive checks (see Appendix figures~\ref{fig:educative_ppc}--\ref{fig:affective_ppc}) show strong alignment between predicted and observed frequencies, indicating good model calibration and fit. The estimated global effects (94\% HDI) across all domains were:

\begin{description}[leftmargin=1.5em, nosep, font=\normalfont\itshape]
\item[Educative:] $\beta_{\mathit{exp}} = 0.21\;[0.12, 0.29];\;
                  \beta_{\mathit{gender}} = -0.25\;[-0.43, -0.06];\;
                  \beta_{\mathit{age}} = -0.08\;[-0.17, 0.00]$.
\item[Explorative:] $\beta_{\mathit{exp}} = 0.36\;[0.27, 0.44];\;
                    \beta_{\mathit{gender}} = -0.42\;[-0.60, -0.25]$.
\item[Affective:] $\beta_{\mathit{exp}} = 0.40\;[0.32, 0.49];\;
                  \beta_{\mathit{gender}} = -0.51\;[-0.69, -0.34]$.
\end{description}

\paragraph{Summary and implications.}  
Across both analytical lenses, CRS experience emerged as the most consistent and positive predictor of interaction-aim importance, supporting its role as a stable input for adaptive user modelling. In contrast, gender effects were notably sparse and domain-specific (e.g., Explorative in \textit{Finance} and Affective in \textit{Housing}), while age showed no reliable influence after multiple-comparison correction. 
These results imply that only CRS experience warrants global policy adaptation, while gender-based adjustments should remain localised and cautious, and age can be excluded as a policy variable. This finding is consistent with the RAE framework’s implementation-agnostic guidance for lean, ethically grounded adaptation in CRS. By deprioritising blunt demographic categories in favour of behavioural signals like system experience, the framework achieves a more robust and personalized user-state representation.

\subsection{User autonomy and dialogue control (H6)}
\label{sec:results-autonomy}

While the preceding analyses established the situational and individual drivers of interactional aims, the practical implementation of the RAE framework requires an understanding of how dialogue initiative should be distributed between the agent and the user. Consequently, we analysed user preferences for dialogue control within the \textit{educative} and \textit{explorative} CRS interaction phases.\footnote{\label{foot:autonomy-exclusion}We did not include a control item for \textit{affective} interactions as affective support is inherently system-led, arising from implicit user cues and contextual inference rather than explicit user direction. Including control ratings for affective behaviour would risk conflating preference with feasibility.}

\paragraph{Descriptive statistics and preference tests.}
Table~\ref{tab:control-desc} reports the descriptive statistics and Wilcoxon signed-rank tests for the two 5-point control items (where 1 = \textit{fully system-initiated} and 5 = \textit{fully user-initiated}). User ratings for both educative and explorative control were significantly above the neutral midpoint (3.0), indicating a clear general preference for user-initiated interaction and providing strong support for H6a. 
The observed effect sizes were medium--large for both educative ($W=1345, z=-5.12, p<.001, r=0.39$) and explorative ($W=1566, z=-4.88, p<.001, r=0.38$) interaction types. Furthermore, a moderate positive correlation between the two items (Spearman's $\rho = 0.47, p<.001$) indicated high internal consistency in autonomy preferences across distinct interactional aims.

\begin{table}[h]
\centering
\caption{Descriptive statistics and one-sample Wilcoxon signed-rank tests for dialogue control preferences ($N=168$). Participants exhibited a significant preference for user-initiated control across both aims ($p < .001$), supporting H6a. Effect sizes ($r$) suggest users value autonomy in directing the conversation during both educative and explorative phases.}
\label{tab:control-desc}
\begin{threeparttable}
\footnotesize 
\setlength{\tabcolsep}{18pt} 
\begin{tabular}{l ccccc}
\toprule
\textbf{Interaction Aim} & \textbf{Mean (SD)} & \textbf{Median} & \textbf{$W$} & \textbf{$z$} & \textbf{$r$} \\
\midrule
Educative   & 3.47 (1.02) & 4.00 & 1345.0 & $-5.12$ & .39 \\
Explorative & 3.45 (0.98) & 3.50 & 1566.0 & $-4.88$ & .38 \\
\bottomrule
\end{tabular}
\begin{tablenotes}
    \scriptsize
    \item \textit{Note:} Scale: 1 = System-initiated, 3 = Neutral, 5 = User-initiated. All tests are compared against the neutral midpoint (3). Spearman's $\rho(\text{Educative, Explorative}) = 0.47, p < .001$.
\end{tablenotes}
\end{threeparttable}
\end{table}

\paragraph{Ordinal logistic regression models.}
To evaluate H6b, we fitted separate ordinal logistic regression models predicting user preferences for control. Predictors included gender (Male=1, Female=0), CRS experience (ordinal), and age (categorical, with 18--24 as the reference group). 
As shown in Table~\ref{tab:control-reg}, gender significantly predicted autonomy preferences within the \textit{educative} model. Specifically, \textit{female} participants exhibited significantly higher odds of preferring user-initiated control (Male coefficient $\beta = -0.79, p = .009$; Female vs. Male OR $= 2.19, 95\%~CI~[1.22, 3.95]$). In contrast, no demographic predictors reached significance in the \textit{explorative} control model. These findings suggest that while overall user preferences favour high autonomy (H6a), demographic modulation is limited to a modest gender-specific effect in educative dialogue (H6b).

\begin{table}[h]
\centering
\caption{Cumulative logit (ordinal) models predicting user-initiated dialogue-control preferences ($N=164$). Notably, Gender emerges as a significant predictor for the \textit{Educative} aim ($p = .009$), where females indicate a significantly higher preference for user-led control compared to males.}
\label{tab:control-reg}
\begin{threeparttable}
\footnotesize 
\setlength{\tabcolsep}{24pt} 
\begin{tabular}{l cc cc}
\toprule
\textbf{Predictor} & \multicolumn{2}{c}{\textbf{Educative}} & \multicolumn{2}{c}{\textbf{Explorative}} \\
\cmidrule(lr){2-3} \cmidrule(lr){4-5}
 & $\beta$ & $p$ & $\beta$ & $p$ \\
\midrule
Gender (Male)   & \textbf{$-0.79$} & \textbf{.009} & $-0.27$ & .349 \\
CRS Experience  & $-0.03$ & .815 & 0.14 & .264 \\
\addlinespace[4pt]
\textit{Age Group}\tnote{a} & & & & \\
25--34 years    & 0.15 & .650 & 0.27 & .420 \\
35--44 years    & $-0.29$ & .489 & 0.09 & .819 \\
45--54 years    & $-0.16$ & .849 & $-0.03$ & .966 \\
55--64 years    & $-0.13$ & .852 & 0.05 & .947 \\
\bottomrule
\end{tabular}
\begin{tablenotes}
    \scriptsize
    \item \textit{Note:} Gender coded as Male=1, Female=0. Negative $\beta$ for Gender indicates higher odds for females to prefer user-initiated control. \tnote{a} Reference group for Age is 18--24 years. Bold indicates statistical significance ($p < .05$).
\end{tablenotes}
\end{threeparttable}
\end{table}

\paragraph{Summary and implications for adaptive CRS design}
Our findings demonstrate a statistically significant preference for user-initiated dialogue control during both educative and explorative exchanges. This autonomy orientation was consistent across interactional aims, supporting H6a. Regarding H6b, gender emerged as the sole demographic predictor, and specifically only within the \textit{educative} context, where females were more likely than males to prefer user-led control.
Within the RAE framework, these autonomy preferences act as a vital user-state signal for dialogue management. While the results do not dictate a specific technical stack, they suggest that adaptive systems can improve user alignment by dynamically adjusting initiative-taking strategies. Given the observed stability of these preferences, implementing a lightweight personalisation rule or model feature can sufficiently support experiential consistency without introducing unnecessary architectural complexity.

\subsection{Summary of hypothesis tests}
\label{sec:hypothesis-summary}

Table~\ref{tab:hypotheses-summary} consolidates all six hypotheses, summarising methods, key findings, and outcomes. Patterns across results indicate that \emph{domain context}, \emph{perceived item value}, \emph{prior CRS experience}, and \emph{autonomy preferences} are the most consistent and discriminative user-state signals for tailoring educative, explorative, and affective interaction aims.
H1–H3 confirmed systematic variation across domains, with high-stakes and cognitively complex contexts elevating educative aims, and affect-rich contexts elevating affective aims. H4 showed that high-value items consistently increased ratings across all aims, underscoring sensitivity to perceived decision stakes. H5 revealed that CRS experience reliably predicted higher ratings, whereas gender effects were rare and domain-specific, and age effects negligible. H6 indicated a general preference for user-initiated dialogue, with only minor demographic modulation.
From an RAE framework perspective, these variables can be modelled as features in an \emph{adaptive CRS state representation}, regardless of underlying implementation. This enables dialogue policies, whether learned or heuristic, to adjust aim weighting, explanation depth, exploratory breadth, and conversational initiative in ways that are context- and user-aligned. The findings thus provide direct, evidence-based parameters for adaptive policy design, extending beyond static interface guidelines toward experience-sensitive conversational strategies.

\begin{table}[h] 
\centering
\begin{threeparttable}
\caption{Summary of hypothesis testing outcomes (H1--H6). 
Findings synthesise frequentist tests (Kruskal--Wallis, Wilcoxon) and Bayesian hierarchical modeling. 
The results confirm that interactional aims are significantly modulated by decision domain (H1--H3) and item value (H4), while user traits (H5) and autonomy preferences (H6) exhibit more nuanced, partial support.}
\label{tab:hypotheses-summary}

\footnotesize 
\begin{tabularx}{\textwidth}{l p{2.8cm} X c}
\toprule
\textbf{Hypothesis} & \textbf{Method(s)} & \textbf{Key Findings} & \textbf{Status} \\
\midrule

H1: Educative / complexity 
& KW, MWU, Bayes 
& Omnibus $H = 93.15, p < .001$; Education $>$ Dining ($r = .36$), Tech $>$ Apparel ($r = .32$). Bayesian $\beta_{\text{Educ.}} = 1.25, 94\%$ HDI $[0.47, 1.96]$. 
& Supported \\
\addlinespace

H2: Explorative / novelty 
& KW, MWU, Bayes 
& Omnibus $H = 33.34, p < .001$; Travel $>$ Wellness ($r = .25$). Bayesian $\beta_{\text{Trav.}} = 1.30, 94\%$ HDI $[0.52, 1.99]$. 
& Supported \\
\addlinespace

H3: Affective / emotion 
& KW, MWU, Bayes 
& Omnibus $H = 25.10, p = .003$; Wellness $>$ Tech ($r = .21$). Bayesian $\beta_{\text{Well.}} = 1.12, 94\%$ HDI $[0.38, 1.83]$. 
& Supported \\
\addlinespace

H4: Item value moderation
& Wilcoxon (paired) 
& High-value items rated significantly higher for all aims ($p < .001$): Educative ($r = .82$); Explorative ($r = .81$); Affective ($r = .86$). 
& Supported \\
\addlinespace

H5: User traits 
& Spearman, Bayes 
& CRS experience positive in 13/30 pairs ($\beta = .21$--$.40$, HDIs $\not\ni 0$). Gender/Age effects largely negligible. 
& Partially Supported \\
\addlinespace

H6: Autonomy / Mode 
& Wilcoxon, Regr. 
& Preference for user-initiated control ($r \approx 0.50, p < .001$). Females prefer user-led educative control (OR $= 2.19, p = .009$). 
& Partially Supported \\

\bottomrule
\end{tabularx}
\begin{tablenotes}
    \scriptsize
    \item \textit{Note:} KW: Kruskal--Wallis; MWU: Mann--Whitney $U$; Bayes: Bayesian hierarchical ordinal regression; HDI: Highest Density Interval. $n=168$. 
    \item Supported: $p < .05$ and Bayesian HDI excludes zero. Partially Supported: Significant effects found in specific subsets only.
\end{tablenotes}
\end{threeparttable}
\end{table}

\section{Discussion}
\label{sec:discussion}

Prior work on CRS has typically examined educative, explorative, or affective aims in isolation, or treated user experience as a static evaluation outcome rather than an adaptive driver of dialogue policy \cite{young2013pomdp, chen2021knowledge, wang2018explainable, yu2019adaptive}. Our findings extend this literature by showing how these aims can be modelled jointly and how their relative importance shifts with domain, item value, and user characteristics. This reframing positions experiential aims not merely as post-hoc evaluation constructs but as actionable state features, providing a bridge between user-centred evaluation frameworks (e.g., ResQue \cite{pu2011user}, CRS-Que \cite{jin2024crs}) and adaptive dialogue management approaches \cite{brabra2021dialogue, chen2018policy}. By demonstrating that educative, explorative, and affective aims operate as dynamic state variables shaped by context and user signals, our study establishes an empirical basis for context-sensitive CRS dialogue management.

\subsection{From empirical evidence to adaptive logic}
\label{sec:discussion-synthesis}

The empirical findings presented in Section~\ref{sec:results} provide more than a statistical validation of our hypotheses; they offer the necessary grounding for a multi-layered approach to conversational adaptation. By quantifying the systematic shifts in user expectations across ten domains, varying stakes, and diverse individual backgrounds, we can move beyond "one-size-fits-all" dialogue policies. The following sections synthesise these insights into the \textit{Recommendation-as-Experience} (RAE) framework, establishing how contextual and user-state signals can be mapped to experience-aligned system behaviours.

\paragraph{\textbf{Domain-specific prioritisation of interaction aims.}}
\label{sec:discussion-domain-moderation}
Results for RQ1 and H1--H3 confirm that the importance users assign to educative, explorative, and affective aims varies systematically across domains. From an adaptive CRS perspective, these patterns can be encoded as \emph{domain-level priors} in the system’s state representation, providing baseline weightings that guide the dialogue policy before finer-grained situational or user-specific signals are incorporated \cite{jannach2021survey, gao2021advances}. Such priors allow a CRS to initialise explanation depth, novelty breadth, and affective tone in ways that match the cognitive and emotional demands characteristic of each application area, thereby supporting context-sensitive interaction from the outset.

\paragraph{Educative aims: cognitive complexity and contextual nuance.}  
As hypothesised in H1, educative aims were highest in complex domains (\textit{Education}, \textit{Technology}), confirmed by a robust Kruskal--Wallis test ($H(9) = 93.15$, $p < .001$). Pairwise effects (e.g., Education vs.\ Dining, $r = 0.36$) show strong differentiation from less cognitively demanding domains. In an adaptive policy model, such results justify higher initial weights for educative behaviours when \texttt{domain\_complexity} is high, triggering richer rationales and scaffolded guidance \cite{zhang2020explainable}. Interestingly, \textit{Travel} also ranked highly for educative aims alongside explorative and affective priorities, suggesting a blended aim configuration. Conversely, \textit{Finance} showed heterogeneous responses, implying sub-domain profiling or user-specific signals are necessary.

\paragraph{Explorative aims: hedonic vs.\ stability-driven contexts.}  
Consistent with H2, explorative aims peaked in novelty-oriented domains (\textit{Travel}, \textit{Entertainment}), $H(9) = 33.34$, $p < .001$. These domains can be tagged with a high \texttt{novelty\_orientation} feature, prompting broader candidate diversity and serendipitous recommendations. Lower explorative prioritisation in \textit{Wellness} illustrates the need for policies that suppress novelty when affective stability is valued, aligning exploration parameters with domain-specific risk profiles \cite{jannach2023evaluating}.

\paragraph{Affective aims: emotional and social salience.}  
Aligned with H3, affective aims dominated in \textit{Wellness} and \textit{Beauty} ($H(9) = 25.10$, $p = .003$), supporting high default weights for empathetic tone and rapport-building behaviours when \texttt{emotional\_salience} is high \cite{raamkumar2022empathetic}. These features can serve as triggers in an adaptive orchestration layer, activating affective templates or tone modulation modules.

These domain-level patterns map directly to policy initialisation parameters (\emph{domain priors}) that can be overridden or reweighted in real time as situational (Section~\ref{sec:discussion-item-value}) and user-specific (Section~\ref{sec:discussion-user-moderation}) signals become available.

\paragraph{\textbf{Item value sensitivity as a dynamic modulator.}}
\label{sec:discussion-item-value}
Addressing H4, perceived item value emerged as a strong within-subject modulator of aim prioritisation. Wilcoxon signed-rank tests confirmed significant increases under high-value scenarios for all three aims: Educative ($r = .82$), Explorative ($r = .81$), and Affective ($r = .86$), with all $p < .001$. These large effects demonstrate that as decision stakes rise, users recalibrate their expectations toward deeper explanation, broader exploration, and richer emotional engagement. This finding resonates with prior research suggesting that decision value amplifies user demands for justification, transparency, and support in recommender interactions~\cite{chen2021values, jin2024crs}.
From an adaptive dialogue management perspective, \texttt{item\_value} serves as a high-sensitivity variable within the state vector, capable of triggering substantial reweighting of aims within a single session. This aligns with the RAE proposition that situational variables can override or amplify baseline policy parameters in real time. \textit{Operational requirements:} Incorporating these insights into functional CRS deployments necessitates that the system treats \texttt{item\_value} as a tracked, dynamic state feature. The architecture must support the mapping of high-value conditions to specific aim-weight adjustments, whether through heuristic thresholds or learned gradients, while ensuring a graceful return to baseline aim levels as situational cues diminish. This mechanism of \emph{layered adaptivity} prevents over-optimisation and user fatigue, ensuring the system remains efficient in low-stakes cases while providing high-fidelity support when stakes are elevated.

\paragraph{\textbf{Moderation by user characteristics.}}
\label{sec:discussion-user-moderation}

Addressing H5, \textit{prior CRS experience} emerged as a consistent predictor of higher ratings across multiple domains. This supports the notion that familiarity with conversational systems fosters higher expectations for depth, novelty, and socially expressive behaviours~\cite{cai2022impacts, jannach2021survey}. In contrast, \textit{gender} effects were narrower and context-contingent, appearing primarily in the \textit{Finance} domain for explorative aims and \textit{Housing} for affective aims. For instance, females assigned significantly higher affective ratings in the \textit{Housing} context (Bayesian $\beta = -0.51$, 94\% HDI excluding zero), suggesting a context-linked sensitivity to socially supportive behaviours~\cite{qazi2022gender}. Conversely, \textit{age} effects were negligible after correction for multiple comparisons, aligning with contemporary findings that user expectations for conversational assistants are more uniformly distributed across age groups than previously theorised~\cite{al2023role}.

\textit{User-centric policy implications.} 
Translating these insights into the RAE framework necessitates a tiered approach to demographic features. Given its predictive robustness, \texttt{crs\_experience} should be treated as a high-confidence behavioural feature for real-time policy adjustment. For expert users, the system should reduce conversational redundancy and increase the injection of novelty; for novices, the policy should shift to prioritise clarity, scaffolding, and interactional reassurance. 
Furthermore, the results suggest that demographic features like \texttt{gender} must be treated as conditional modulators, applied only in specific domains where reproducible, context-specific effects have been validated. Meanwhile, \texttt{age} can be deprioritised as a primary policy variable due to its negligible predictive utility in this context. This layered treatment prevents demographic features from acting as blunt segmentation tools, instead applying them as targeted modulators only when empirical evidence supports their inclusion. 
Within the RAE framework, user traits thus operate as \emph{secondary modulators} that shape the baseline domain--value policy blend. This conservative application of demographics ensures the system remains adaptive without falling into the trap of over-segmentation. For a detailed discussion on the ethical considerations of this approach, including safeguards against stereotyping, see~\ref{sec:responsible-adaptation}.

\paragraph{\textbf{User autonomy and dialogue control preferences.}}
\label{sec:discussion-autonomy}

Addressing H6a and H6b, participants showed a clear preference for \emph{user-initiated} dialogue for both the \textit{educative} (M = 3.47; median = 4) and \textit{explorative} (M = 3.45; median = 3.5) aims. One-sample Wilcoxon signed-rank tests against the neutral midpoint ($3 =$ equal user/system initiative) indicated ratings above neutral for both aims (Educative: $W=1345$, $n'=110$, $p<.001$, $r=0.51$, rank-biserial $=0.56$, CLES $=0.78$; Explorative: $W=1566$, $n'=116$, $p<.001$, $r=0.49$, rank-biserial $=0.54$, CLES $=0.77$; Table~\ref{tab:control-desc}). A moderate positive association between the two aims (Spearman's $\rho=0.471$, $p<.001$) suggests that autonomy preferences are broadly consistent across educative and explorative contexts.\footnote{$n'$ denotes the number of non-tied observations (ratings $\neq 3$) used by the Wilcoxon test.} These findings echo broader work in conversational systems, where users frequently prefer to retain initiative in information-rich interactions \cite{laranjeiro2015survey, tsai2021effects, kocaballi2019personalization}.

\paragraph{Demographic moderation.}
Ordinal logistic regression identified \emph{gender} as the only reliable demographic predictor, and only for the \textit{educative} aim: \emph{female} participants had higher odds of preferring user-initiated control (Male coefficient $\,\beta=-0.785$, $p=.009$; Female vs.\ Male OR $=2.19$, 95\% CI [1.22, 3.95]; see Table~\ref{tab:control-reg}). Age and CRS experience were non-significant. Notably, domain-specific gender effects observed for aim \emph{importance} (e.g., explorative--Finance; affective--Housing) did not map onto corresponding differences in \emph{control} preference; the detected control effect appears specific to educative dialogue.

\paragraph{Context note.}
The dialogue-control items were framed within an apparel e-commerce context (clothing/accessories/footwear): a relatively low-to-moderate stakes, frequent decision setting. Generalisation to other CRS domains should therefore consider this contextual grounding.

\paragraph{Policy implications.}
Within the RAE framework, \\ \texttt{autonomy\_pref} can be modelled as a state feature that allocates conversational initiative:
\begin{description}[leftmargin=*, nosep]
  \item High \texttt{autonomy\_pref} $\rightarrow$ system defers to user-initiated turns; unsolicited prompts are minimised except under high-stakes conditions or when clarification is urgent.
  \item Low \texttt{autonomy\_pref} $\rightarrow$ system adopts more initiative, proactively scaffolding the dialogue while maintaining easy opt-out.
\end{description}

For affective behaviours, often more effective when timely and unsolicited, system-initiated turns can remain the default, triggered by sentiment cues and always overrideable by the user \cite{yang2021designing, li2025systematics}.

\paragraph{Interaction with domain and value.}
Autonomy preferences interact with domain profile and perceived value: in educative-dominant, higher-value contexts (e.g., \textit{Education}, \textit{Technology}), user-initiated turns can enhance trust and comprehension; in affect-rich contexts (e.g., \textit{Wellness}, \textit{Beauty}), gentle system-initiated affective support may better sustain rapport. Thus, \texttt{autonomy\_pref} complements domain profile, item value, and user traits to yield an adaptive initiative strategy within the RAE state--policy mapping.

\subsection{RAE perspectives and framework}
\label{sec:discussion-rae}

Findings from H1--H6 show that the perceived importance of educative, explorative, and affective aims is shaped by layered contextual and individual factors that change within and across interactions.

\textit{Dominant contextual drivers.}
Domain and item value jointly explained markedly more variance in aim ratings (Bayesian $R^2 = 0.39$) than user characteristics alone ($R^2 = 0.07$). Domain effects were robust across all aims (e.g., Educative: $H = 93.15$, $p < .001$), with medium contrasts such as \textit{Education} $>$ \textit{Dining} ($r = .36$) and \textit{Travel} $>$ \textit{Wellness} ($r = .25$) (Table~\ref{tab:kruskal-wallis-results}). Item value produced strong within-subject shifts for all aims (Wilcoxon $r = .82$ to $r = .86$, $p < .001$; Table~\ref{tab:h4-wilcoxon}), confirming prior observations that decision stakes amplify expectations for explanation and support \cite{chen2021values, wang2018explainable, jin2024crs}.

\textit{Secondary individual differences.}
CRS experience predicted higher ratings in 13/30 domain--aim pairs ($\rho \approx .20$; Bayesian $\beta = 0.21$--$0.40$), consistent with prior work showing that familiarity fosters higher expectations for depth, novelty, and socially expressive behaviours \cite{cai2022impacts, nourani2020role}. Gender effects were narrower and context-specific, echoing patterns reported in prior studies on interaction style preferences \cite{qazi2022gender, rana2021effect}, while age was negligible (Table~\ref{tab:demographic-correlations}). Autonomy preferences favoured user-initiated control for educative and explorative aims (Wilcoxon $r = .70$ and $r = .68$, both $p < .001$), aligning with evidence that mismatches in initiative reduce trust and usability \cite{del2024ai, kraus2021modelling, bonicalzi2023artificial}, though participants tolerated system initiative when scaffolding complex tasks or providing timely affective support.

\textit{Implication.}
These findings substantiate the RAE proposition that adaptivity in CRS should be layered: domain and value provide the primary baseline, user traits act as secondary refinements when predictive, and autonomy preferences allocate initiative. We formalise this as a state--policy mapping in the next subsection \ref{sec:rae-adaptation}.

\paragraph{The RAE adaptation framework.}
\label{sec:rae-adaptation}

The \emph{Recommendation-as-Experience (RAE)} adaptation framework translates the synthesis into a layered control model for CRS. The dialogue manager maintains a \emph{state vector} with four feature groups: 
(1) \texttt{domain\_profile} (task complexity, novelty orientation, emotional salience), 
(2) \texttt{item\_value} (stakes), 
(3) \texttt{user\_traits} (behavioural history, conditional demographics), and 
(4) \texttt{autonomy\_pref} (initiative pattern). 
A policy function maps this state to an \emph{aim-weight vector} over educative, explorative, and affective aims. Formally, at any dialogue turn $t$, the conversational state is represented as the vector:
$$\mathbf{s}_t = \langle \mathcal{D}, \mathcal{V}_t, \mathcal{U}_{\mathit{exp}}, \mathcal{U}_{\mathit{aut}}, \mathbf{H}_t \rangle$$
where $\mathcal{D}$ represents the domain priors, $\mathcal{V}_t$ the dynamic item value, $\mathcal{U}_{\mathit{exp}}$ and $\mathcal{U}_{\mathit{aut}}$ the persistent user traits and autonomy preferences, and $\mathbf{H}_t$ the latent dialogue history. The adaptation policy $\pi$ then maps this state to the weight vector $\mathbf{a}_t$:
$$\pi(\mathbf{s}_t) \rightarrow \mathbf{a}_t, \quad \text{where } \mathbf{a}_t = \{w_{\mathit{edu}}, w_{\mathit{exp}}, w_{\mathit{aff}}\}$$

\begin{figure}[]
\centering
\begin{tikzpicture}[
    node distance=0.8cm and 0.8cm, 
    box/.style={
        rectangle, draw, thick, rounded corners, fill=white, 
        text width=3cm, 
        minimum height=1.1cm, 
        align=center, 
        font=\scriptsize,
        inner sep=3pt 
    },
    detail/.style={
        rectangle, draw, thin, dashed, fill=blue!5, 
        text width=2.2cm, 
        minimum height=0.8cm, 
        align=center, 
        font=\scriptsize,
        inner sep=2pt
    },
    arrow/.style={-Stealth, thick},
    label/.style={font=\tiny\itshape, align=center}
]

\node (q1) [box] {\textbf{Stage 1: Elicitation} \\ Scenario-based Survey \\ ($N=168$)};

\node (q2) [box, right=of q1] {\textbf{Stage 2: Analysis} \\ Bayesian Models \\ \& Effect Sizes ($r$)};

\node (q3) [box, right=of q2] {\textbf{Stage 3: RAE Framework} \\ State Vector $\mathbf{s}_t$ \\ \& Policy $\pi$};

\draw [arrow] (q1) -- node[above, font=\tiny] {Data} (q2);
\draw [arrow] (q2) -- node[above, font=\tiny] {Priors} (q3);

\node (b1) [detail, below=of q1] {High vs. Low \\ Value Vignettes};
\node (b2) [detail, below=of q2] {Empirical Weight \\ Validation};
\node (b3) [detail, below=of q3] {Adaptive Policy \\ Mapping};

\draw [thin, gray] (q1) -- (b1);
\draw [thin, gray] (q2) -- (b2);
\draw [thin, gray] (q3) -- (b3);

\end{tikzpicture}
\caption{The three-stage methodological workflow for operationalising the Recommendation-as-Experience (RAE) framework. Stage 1 elicits user preferences through high-stakes vignettes; Stage 2 utilises Bayesian modeling to derive robust empirical weights and effect sizes ($r$); and Stage 3 translates these weights into a formal computational policy $\pi$ and state vector $\mathbf{s}_t$ for adaptive system behaviour. Dashed boxes represent the specific sub-components validated at each phase of the pipeline.}
\label{fig:methodology-mapping}
\end{figure}
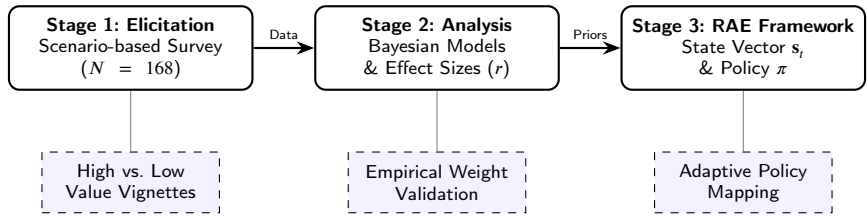

The transition from empirical user expectations to computational system parameters is illustrated in Figure~\ref{fig:methodology-mapping}, which maps the significant effect sizes derived from our multi-domain study to the primary weights of the RAE state vector.
This mapping is \emph{implementation-agnostic} and can be realised as rules, learned policies, or LLM-based orchestration layers \cite{young2013pomdp, xu2024leveraging, kwan2023survey, jannach2021survey, xu2024leveraging}. Table~\ref{tab:rae-adaptation} encodes rules consistent with our post-hoc analyses; these serve as defaults that a learned or hybrid policy can refine. To illustrate how these rules manifest in real CRS use, we now provide design vignettes.

\begin{table}[t]
\centering
\begin{threeparttable}
\caption{Context-sensitive modulation of interactional aims within the Recommendation-as-Experience (RAE) framework. 
The table illustrates how the system dynamically shifts its emphasis across Educative (Edu.), Explorative (Exp.), and Affective (Aff.) dimensions based on domain-specific triggers and situational complexity. 
By mapping these high-level clusters to preferred dialogue modes (e.g., system-led vs. user-led), the framework facilitates an adaptive initiative-taking policy that balances epistemic decision support with affective alignment.}
\label{tab:rae-adaptation}

\scriptsize
\setlength{\tabcolsep}{0pt}
\begin{tabular*}{\textwidth}{@{\extracolsep{\fill}} p{3cm} ccc p{2.5cm} p{3.5cm}}
\toprule
\textbf{Domain cluster} & \textbf{Edu.} & \textbf{Exp.} & \textbf{Aff.} & \textbf{Preferred mode} & \textbf{Indicative triggers}\\
\midrule
High-stakes / Complex & \ccmark & -- & -- & System $\rightarrow$ User & High item value; extreme decision complexity \\
\addlinespace
Cross-cutting (e.g., Travel) & \cmark & \cmark & \cmark & Mixed-initiative & Multi-criteria trade-offs; novelty seeking \\
\addlinespace
Hedonic / Leisure & -- & \ccmark & \cmark & User-led or Mixed & Visual novelty cues; trend-based browsing \\
\addlinespace
Affect-rich / Identity & \cmark & \cmark & \ccmark & Gentle System Init. & Emotional salience; expressed sentiment \\
\addlinespace
Social / Contextual & \cmark & \ccmark & \cmark & User-led / Nudges & Occasion framing; group decision context \\
\addlinespace
Functional / Pragmatic & \ccmark & -- & -- & System $\rightarrow$ User & Specification comparison; expertise signals \\
\bottomrule
\end{tabular*}

\begin{tablenotes}
    \scriptsize
    \item \textit{Note:} \ccmark: Primary/Strong emphasis; \cmark: Secondary/Moderate emphasis; --: Low/De-emphasized. 
    \item Preferred dialogue mode specifies the allocation of conversational initiative based on the predicted user information need.
\end{tablenotes}
\end{threeparttable}
\end{table}

\FloatBarrier

\subsubsection{Illustrative design vignettes and rule-set}

Building directly on the synthesis presented in Section~\ref{sec:discussion-rae}, we illustrate how the layered predictors of domain, item value, user traits, and autonomy preferences are operationalised through the RAE framework. To demonstrate the framework's flexibility, we present two contrasting interaction scenarios.

\textit{Scenario A: High-Stakes Travel Planning.} 
Consider a user planning a once-in-a-lifetime honeymoon, representing a \textit{high-value} situational context within the \textit{Travel} domain. In this scenario, the \textit{domain layer} is characterised as novelty-heavy and moderately complex; consequently, the RAE framework assigns balanced emphasis to explorative and educative aims (Section~\ref{sec:discussion-domain-moderation}). Within the \textit{user layer}, we observe a highly experienced CRS user, triggering a shift in system tone toward concise and socially engaging responses (Section~\ref{sec:discussion-user-moderation}). Finally, at the \textit{situational layer}, the high \texttt{item\_value} signals the need for deeper explanations, wider option diversity, and the activation of affect-sensitive phrasing (Section~\ref{sec:discussion-item-value}). The resulting interaction surfaces serendipitous travel options paired with transparent trade-offs, such as detailed cost breakdowns and contextualised justifications.

\textit{Scenario B: Postgraduate Education Selection.} 
In contrast, consider a user evaluating postgraduate study options. Here, the \textit{Education} domain profile is primarily cognitively demanding and high-stakes. At the \textit{domain layer}, educative aims dominate to provide the structured comparisons and rationale necessary for such a significant decision. Within the \textit{user layer}, the system identifies a novice user with limited CRS experience; the dialogue policy thus prioritises scaffolding, clarity, and step-by-step guidance. Given the high \texttt{item\_value} at the \textit{situational layer}, the system proactively surfaces additional resources such as career outcomes and alumni reviews to bolster user confidence. The system ultimately highlights degree options with transparent trade-offs (e.g., tuition fees vs. global ranking), carefully balancing the depth of functional explanation with active reassurance.

\textit{Operationalising the RAE Framework.} 
These narratives illustrate how domain profiles, user traits, and situational value cues jointly determine adaptive aim blends. This mapping logic can be realised through various computational architectures, ranging from heuristic rule-sets to learned policies or LLM-based orchestration layers~\cite{young2013pomdp, zhu2025recommender}. As demonstrated in the pseudocode below, the RAE state vector serves as a principled input for these policies, ensuring that real-time system behaviour remains grounded in validated predictors.

\begin{adjustwidth}{1.5em}{1.5em}
\begin{lstlisting}[language=Python, basicstyle=\footnotesize\ttfamily, frame=none, breaklines=true]
# State vector derived from multi-source signals
state = {
    "domain_profile": "Travel", 
    "item_value": "High",
    "user_traits": {"crs_experience": "High"},
    "autonomy_pref": "User-led"
}

# (a) Rule-based Mapping (Heuristic Policy)
if state["item_value"] == "High":
    set_aim_weights(educative=0.8, explorative=0.6, affective=0.7)

# (b) Learned Policy (Supervised/RL Mapping)
# Vectorizes state and predicts weight distribution
weights = policy_model.predict(encode(state))

# (c) Orchestration (LLM adaptation)
# Adapts prompt instructions based on aim weights
prompt = adapt_prompt(base_template, weights)
response = generator(prompt)
\end{lstlisting}
\end{adjustwidth}

This state-driven approach ensures that the RAE framework remains independent of the specific implementation stack while providing a conceptual vocabulary for designing context-sensitive recommendations.

\subsubsection{Design guidance}
The RAE framework facilitates adaptation across three primary dimensions:

\textit{System initialisation.} 
Levers include domain-aware defaults, where weights are derived from a \texttt{domain\_profile}, and aim-aligned evaluation, which complements standard ranking metrics with specific experiential measures \cite{jin2024crs}.

\textit{User-centric personalisation.} 
Personalisation is achieved through behavioural calibration based on CRS experience and configurable autonomy, allowing users to manage initiative through interface elements like detail--discovery sliders.

\textit{Contextual dynamics.} 
The system maintains relevance via real-time value triggers that scale aims during high-stakes episodes and uses transparent adaptivity to communicate these transitions to the user.

These findings establish a principled foundation for encoding experiential aims as adaptive state variables in CRS. Building on established foundations of explainability \cite{Berkovsky2016, zhang2020explainable}, exploratory search \cite{marchionini2006exploratory, chen2021exploration}, and affective support \cite{sneha2024affective, djamasbi2010affect, raamkumar2022empathetic}, our study addresses a critical gap in the literature.
While prior research has significantly advanced ranking accuracy and intent modelling \cite{yu2019adaptive, vakulenko2017conversational}, as well as conversational fluency and empathy \cite{wardatzky2025whom, raamkumar2022empathetic}, these experiential strands have largely been studied in isolation or treated as static evaluation outcomes. Our study addresses this gap by providing empirical evidence for \textit{when} and \textit{how} educative, explorative, and affective aims should be foregrounded jointly.
Building on these insights, the proposed RAE adaptation framework \emph{conceptually outlines} how contextual signals (domain, value, user traits) are mapped to policy mechanisms and to specific behavioural realisations: \textit{educative} (justifications, contrastive comparisons), \textit{explorative} (serendipitous suggestions, novelty framing), and \textit{affective} (empathetic tone, sentiment alignment). By treating these aims as adaptive state variables, the RAE perspective provides both a conceptual vocabulary and practical levers for designing personalised, transparent, and context-sensitive conversational recommendations.

\paragraph{Responsible adaptation and safeguards.}
\label{sec:responsible-adaptation}
In applying the RAE framework, we recommend a conservative stance on personalisation. \emph{Behavioural signals} (e.g., prior CRS experience) should be prioritised as primary drivers of adaptation, while \emph{demographic variables} are treated only as conditional features, used selectively and only where validated, context-specific effects exist. This approach prevents demographic attributes from serving as crude segmentation criteria and mitigates risks of stereotyping or unfair treatment \cite{friedman2022excerpt, mehrabi2021survey, chen2021values}. In addition, policy shifts should be logged for auditability and user oversight, ensuring that adaptive behaviours remain transparent, accountable, and respectful of user autonomy.

\subsection{Limitations and future work}

This study’s cross-sectional design limits observation of evolving goals and behaviours in real-world CRS use. Longitudinal and field studies are needed to assess the ecological validity and stability of the RAE adaptation framework.
Our modelling relied on Bayesian hierarchical ordinal regression; future work should test alternative approaches and integrate richer behavioural signals, including sequential interaction data.
Demographic measures were coarse. Incorporating psychological and cognitive constructs may deepen understanding of user heterogeneity.
Autonomy preferences were examined in e-commerce context; multi-domain studies could reveal broader patterns relevant to adaptive dialogue management.
The framework is implementation-agnostic and can be operationalised in LLM-based, rule-based, or hybrid CRS architectures. As it is derived from a single dataset, iterative validation through longitudinal deployments, A/B testing of policy rules or learned controllers, and integration of multimodal sensing and reinforcement or preference learning will be essential to realise fully adaptive, context-aware CRS.

\section{Conclusion}
\label{sec:conclusion}

This study advances the design of user-centred CRS through the formalisation of the \emph{Recommendation-as-Experi-\allowbreak ence} (RAE) adaptation framework. By conceptualising recommendation as an unfolding, situated interaction rather than solely a prediction task, we provide an architecture-agnostic blueprint for aligning system behaviours with experiential quality. Empirically grounded in a multi-domain study ($N=168$), our results quantify how users jointly prioritise three core interaction aims: \textit{educative}, \textit{explorative}, and \textit{affective}. 
We demonstrate that these priorities are systematically modulated by domain context, perceived item value, and user-level traits. Educative aims dominate in high-stakes, cognitively complex domains such as housing and finance, while affective aims gain prominence in emotionally salient contexts such as wellness. Furthermore, perceived item value acts as a critical dynamic modulator, significantly elevating expectations across all interaction aims in high-stakes scenarios. While demographic influences appeared modest, prior CRS experience and stable autonomy preferences emerged as robust signals for adaptive user modelling.
The RAE framework operationalises these insights through a layered \emph{state--policy--behaviour} mapping. By encoding contextual and user signals into structured state representations, the framework enables dialogue policies to adaptively modulate explanation depth, retrieval diversification, and affective tone. This shift from performance-centric optimisation to holistic experiential alignment allows for seamless integration with rule-based, hybrid, or LLM-based controllable generation. 
Future work should validate RAE through longitudinal field deployments to assess the stability of these experiential priorities over time. Additionally, exploring the integration of RAE state representations with adaptive policy learning or automated prompt-orchestration mechanisms will further strengthen its ability to deliver context-aware, experientially rich conversational recommendation in practice.

\section*{Acknowledgements}

During the preparation of this work the authors used ChatGPT 5 in order to improve readability and language of the final draft. After using this tool/service, the authors reviewed and edited the content as needed and take full responsibility for the content of the published article.

\FloatBarrier

\appendix


\section{Interaction preferences across domains}
\setcounter{figure}{0}
\renewcommand{\thefigure}{A.\arabic{figure}}

\begin{figure}
  \centering
  \includegraphics[width=0.9\textwidth]{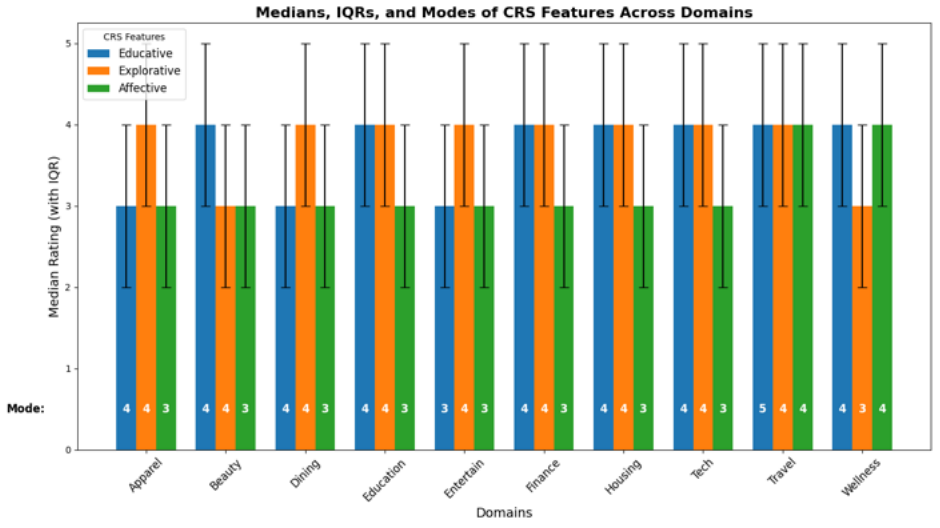}
  \caption{Descriptive distribution of ratings across domains. Medians (markers), modes (labels), and interquartile ranges (error bars) illustrate central tendency and variability for educative, explorative, and affective interaction. These statistics contextualise the domain-level effects reported in Section~\ref{sec:results}.}
  \label{fig:stacked_bar}
\end{figure}

To evaluate H1--H3, we tested whether perceived importance ratings for educative, explorative, and affective interaction varied across application domains. Participants rated each strategy on a 5-point Likert scale for each domain. Figure~\ref{fig:stacked_bar} reports descriptive statistics (medians, modes, interquartile ranges), complementing the inferential analyses in Section~\ref{sec:results}.

\section{Bayesian posterior predictive checks}
\setcounter{figure}{0}
\renewcommand{\thefigure}{B.\arabic{figure}}

\begin{figure}
  \centering
  \includegraphics[width=0.7\textwidth]{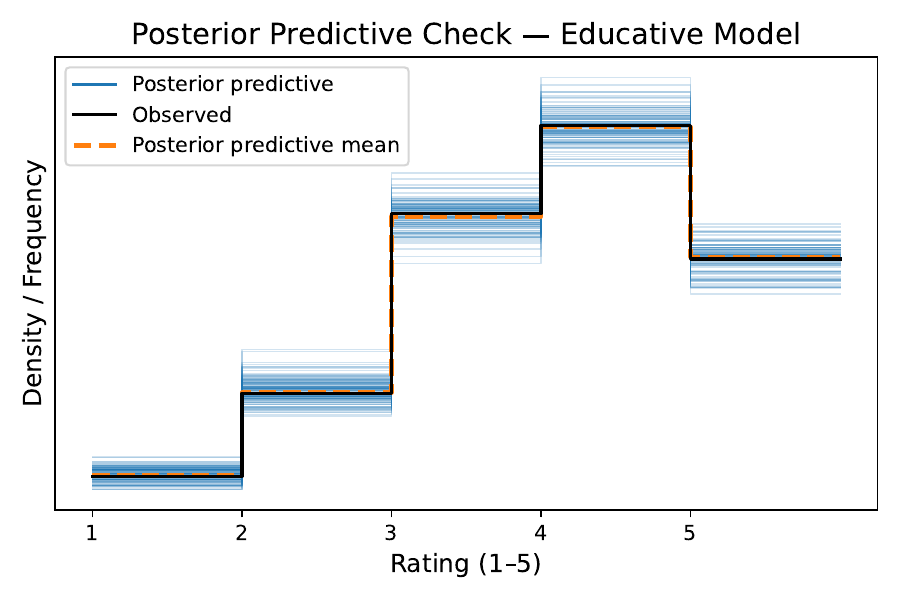}
  \caption{Posterior predictive check for the \textbf{Educative} model. Predicted distributions closely align with observed responses across all ordinal categories, indicating robust fit.}
  \label{fig:educative_ppc}
\end{figure}

\begin{figure}
  \centering
  \includegraphics[width=0.7\textwidth]{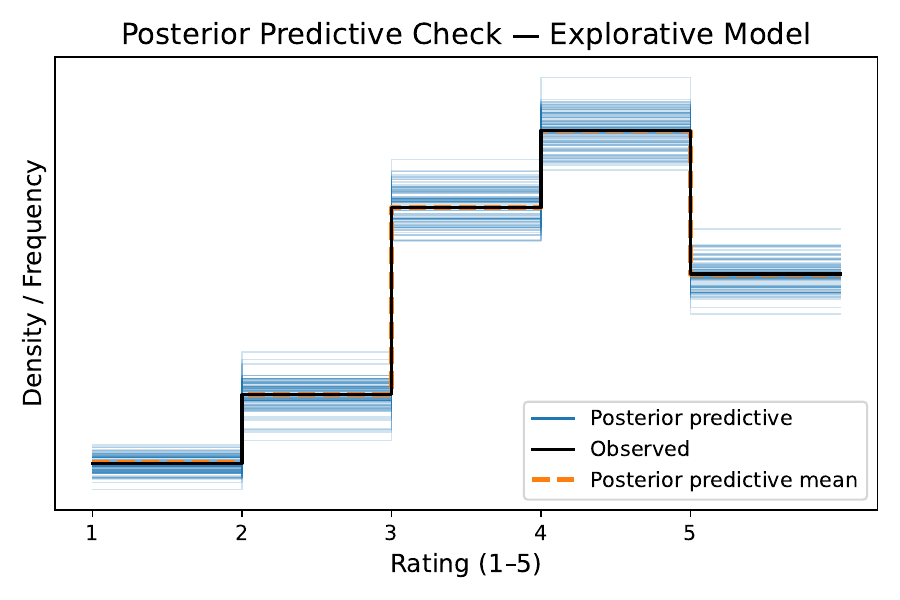}
  \caption{Posterior predictive check for the \textbf{Explorative} model. Predicted and observed distributions exhibit strong agreement, supporting model adequacy.}
  \label{fig:explorative_ppc}
\end{figure}

\begin{figure}
  \centering
  \includegraphics[width=0.7\textwidth]{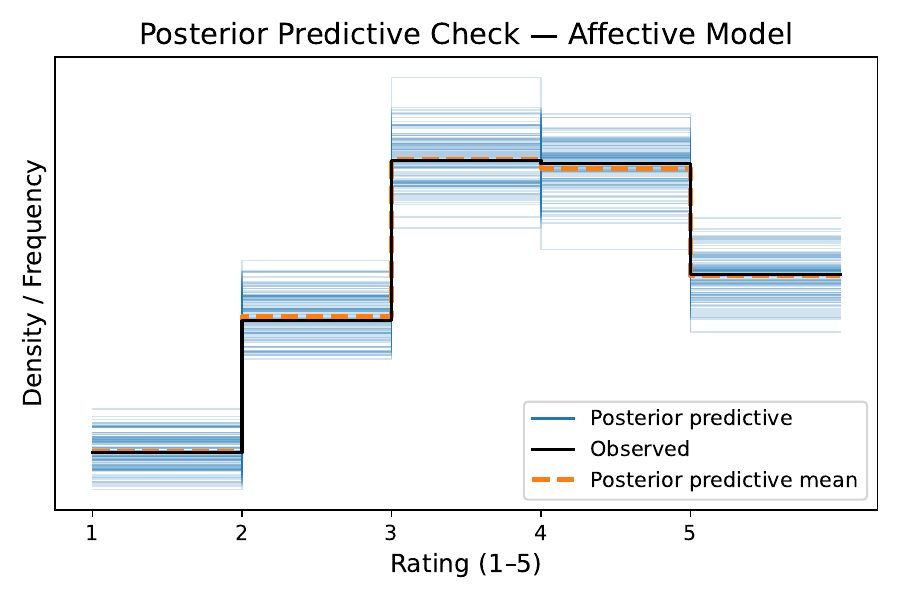}
  \caption{Posterior predictive check for the \textbf{Affective} model. Predicted distributions align well with observed data, providing evidence of model validity.}
  \label{fig:affective_ppc}
\end{figure}

Posterior predictive checks (PPCs) were conducted for each model to evaluate whether simulated responses replicated the observed Likert-scale distributions (Figures~\ref{fig:educative_ppc}--\ref{fig:affective_ppc}). In each plot, shaded densities show predicted ordinal response distributions, black lines indicate observed values, and dashed lines mark predictive means. Close alignment between predicted and observed distributions provides evidence of satisfactory model fit. Full Bayesian regression results are reported in Section~\ref{sec:results-bayesian-models}.

\section{RAE adaptation framework (triangle view)}
\setcounter{figure}{0}
\renewcommand{\thefigure}{C.\arabic{figure}}

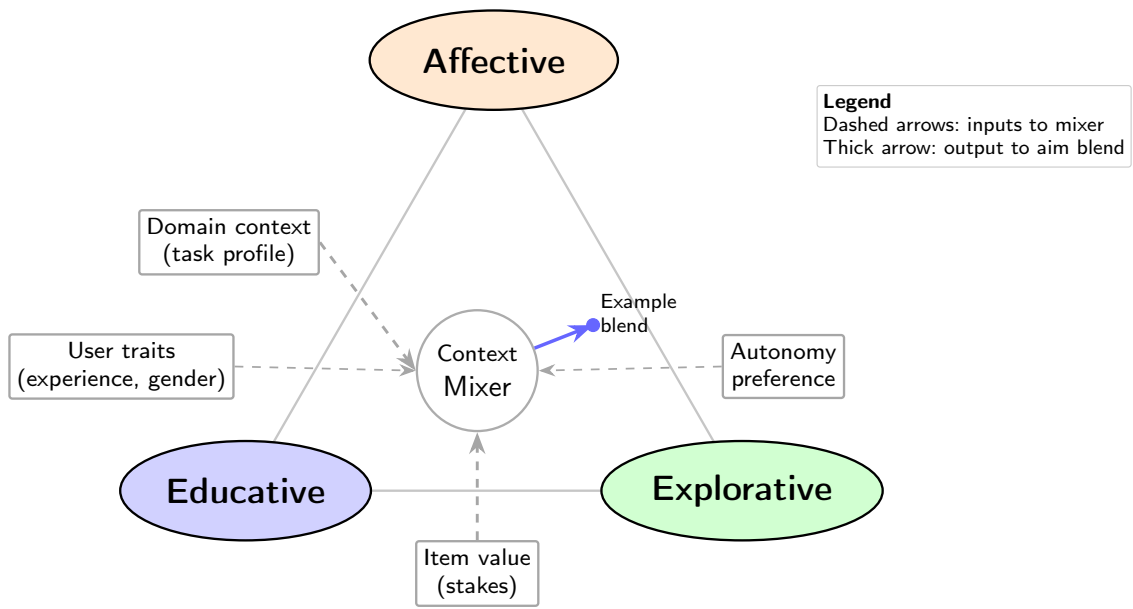
\begin{figure}
  \centering
  \resizebox{0.9\linewidth}{!}{%
  \begin{tikzpicture}[
      font=\normalsize,
      aim/.style={ellipse, draw, thick, fill=#1!18, minimum width=3.0cm, minimum height=1.2cm, font=\large\bfseries, align=center},
      hub/.style={circle, draw=gray!65, thick, fill=white, minimum size=10mm, align=center},
      inbox/.style={draw=gray!70, rounded corners=1pt, thick, fill=white, align=center, font=\footnotesize, inner sep=2.5pt},
      inarrow/.style={->, dashed, gray!70, line width=0.6pt, >=Stealth},
      outarrow/.style={->, very thick, blue!60, >=Stealth}
    ]
    \coordinate (A) at (0,0);
    \coordinate (B) at (6,0);
    \coordinate (C) at (3,5.2);
    \draw[thick, gray!45] (A)--(B)--(C)--cycle;
    \node[aim=blue]   at (A) {Educative};
    \node[aim=green]  at (B) {Explorative};
    \node[aim=orange] at (C) {Affective};
    \coordinate (Cent) at (3,1.73);
    \fill[black] (Cent) circle (2pt);
    \node[hub] (hub) at (2.8,1.45) {\footnotesize Context\\Mixer};
    \coordinate (State) at (4.2,2.0);
    \draw[outarrow] (hub) -- (State);
    \filldraw[blue!60] (State) circle (2.2pt);
    \node[font=\scriptsize, align=left] at (4.75,2.15) {Example\\blend};
    \node[inbox] (domain) at (-0.2,3) {Domain context\\(task profile)};
    \node[inbox] (user)   at (-1.5,1.5) {User traits\\(experience, gender)};
    \node[inbox] (value)  at (2.8,-1.0) {Item value\\(stakes)};
    \node[inbox] (auto)   at (6.5,1.5) {Autonomy\\preference};
    \draw[inarrow, line width=1.0pt] (domain.east) -- (hub.west);
    \draw[inarrow]                    (user.east)   -- (hub.west);
    \draw[inarrow, line width=1.0pt] (value.north)  -- (hub.south);
    \draw[inarrow]                    (auto.west)   -- (hub.east);
    \node[draw=gray!40, fill=white, rounded corners=1pt, inner sep=2pt, font=\scriptsize, anchor=north west] at (6.9,4.9) {%
      \begin{tabular}{@{}l@{}}
        \textbf{Legend}\\
        Dashed arrows: inputs to mixer\\
        Thick arrow: output to aim blend
      \end{tabular}};
  \end{tikzpicture}}
  \caption{RAE adaptation framework (triangle view). Educative, explorative, and affective aims form the vertices of a ternary space. Contextual inputs (domain, user traits, item value, autonomy preference) feed into a mixer, yielding an adaptive blend of aims that guides CRS behaviour.}
  \label{fig:rae-framework}
\end{figure}

For clarity, Figure~\ref{fig:rae-framework} presents the RAE adaptation framework in a triangular view. The three interaction aims define a ternary space, with contextual signals feeding into a mixer that produces a blended behavioural orientation. This schematic supplements the architectural description in Section~\ref{sec:rae-adaptation}.

\clearpage
\bibliographystyle{model1-num-names}

\bibliography{cas-refs}


\end{document}